%% file: manuscript.tex
\documentclass[journal]{IEEEtran} 
\usepackage[bookmarks=false]{hyperref}
\hypersetup{
    colorlinks,
    citecolor=black,
    filecolor=black,
    linkcolor=black,
    urlcolor=black
}

\ifCLASSINFOpdf
  \usepackage[pdftex]{graphicx}
  \usepackage{float}
  \graphicspath{{figures/}}
  \DeclareGraphicsExtensions{.pdf,.jpeg,.png}
\else
  \usepackage[dvips]{graphicx}
  \DeclareGraphicsExtensions{.pdf,.jpeg,.png,.eps}
\fi

\usepackage{amsmath, amsfonts, mathtools, empheq}
\usepackage{amssymb}  
\usepackage[table, dvipsnames]{xcolor}
\usepackage{enumitem}
\usepackage{comment}
\definecolor{color1}{HTML}{F2F3F4}
\definecolor{color2}{HTML}{F0F8FF}
\definecolor{color3}{HTML}{F0FFFF}
\definecolor{myyellow}{HTML}{EAECEE}

\usepackage{lipsum,adjustbox}
\usepackage{pbox}
\usepackage[caption=false,font=normalsize,labelfont=sf,textfont=sf]{subfig}

\usepackage{multirow}
\usepackage{longtable}
\usepackage{colortbl}
\usepackage{rotating}
\usepackage{cite}
\usepackage{xurl}
\usepackage{hyperref}
\usepackage{booktabs}
\usepackage{metalogo}
\usepackage{tabularx}
\usepackage{tabulary}
\usepackage{array}

\usepackage{balance}
\usepackage{xtab, afterpage}

\usepackage [english]{babel}
\usepackage [autostyle, english = american]{csquotes}
\MakeOuterQuote{"}

\usepackage{bigstrut}
\usepackage{array}
\newcolumntype{L}{>{$}l<{$}}
\newcolumntype{C}{>{$}c<{$}}
\newcolumntype{R}{>{$}r<{$}}

\usepackage{forest,array}
\usepackage{tikz}
\usetikzlibrary{shapes,shadows,arrows}
\usepackage{changepage,threeparttable} 
\usepackage{pifont}
\newcommand{\xmark}{\ding{55}}%

\newcommand{\etal}{\textit{et al.}}
\newcommand*\rot{\rotatebox{90}}
\newcommand{\emna}[1]{\textcolor{black}{#1}}
\newcommand{\zina}[1]{\textcolor{black}{#1}}
\newcommand{\asif}[1]{\textcolor{black}{#1}}

\newcommand{\khan}[1]{\textcolor{black}{#1}}
\newcommand{\eb}[1]{\textcolor{black}{#1}}
\newcommand{\commentout}[1]{}

\usepackage{colortbl}
\begin{document}
%
\title{A Survey on Mobile Edge Computing for Video Streaming: Opportunities and Challenges}
%
%
%

\author{
  \IEEEauthorblockN{
  	Muhammad Asif Khan\IEEEauthorrefmark{1},
  	Emna Baccour\IEEEauthorrefmark{3},
    Zina Chkirbene\IEEEauthorrefmark{2},
    Aiman Erbad\IEEEauthorrefmark{3},
    Ridha Hamila\IEEEauthorrefmark{2},
    Mounir Hamdi\IEEEauthorrefmark{3} and
    Moncef Gabbouj\IEEEauthorrefmark{4}
  }

  \IEEEauthorblockA{
  	Qatar University\IEEEauthorrefmark{1}\IEEEauthorrefmark{2},
    Hamad Bin Khalifa University\IEEEauthorrefmark{3},
    Tampere University\IEEEauthorrefmark{4} \\
    Email:
        \{asifk, aerbad\}@ieee.org\IEEEauthorrefmark{1}\IEEEauthorrefmark{3}, 
        \{hamila, zina.chk\}@qu.edu.qa\IEEEauthorrefmark{2}, 
        \{ebaccourepbesaid, mhamdi\}@hbku.edu.qa\IEEEauthorrefmark{3},
        moncef.gabbouj@tuni.fi\IEEEauthorrefmark{4} 
  }
  }


\maketitle

\begin{abstract}

5G communication brings substantial improvements in the quality of service provided to various applications by achieving higher throughput and lower latency. However, interactive multimedia applications (e.g., ultra high definition video conferencing, 3D and multiview video streaming, crowd-sourced video streaming, cloud gaming, virtual and augmented reality) are becoming more ambitious with high volume and low latency video streams putting strict demands on the already congested networks. Mobile Edge Computing (MEC) is an emerging paradigm that extends cloud computing capabilities to  the edge of the network i.e., at the base station level. To meet the latency requirements and avoid the end-to-end communication with remote cloud data centers, MEC allows to store and process video content (e.g., caching, transcoding, pre-processing) at the base stations. Both video on demand and live video streaming can utilize MEC to improve existing services and develop novel use cases, such as video analytics, and targeted advertisements. MEC is expected to reshape the future of video streaming by providing ultra-reliable and low latency streaming (e.g., in augmented reality, virtual reality, and autonomous vehicles), pervasive computing (e.g., in real-time video analytics), and blockchain-enabled architecture for secure live streaming. This paper presents a comprehensive survey of recent developments in MEC-enabled video streaming bringing unprecedented improvement to enable novel use cases. A detailed review of the state-of-the-art is presented covering novel caching schemes, optimal computation offloading, cooperative caching and offloading and the use of artificial intelligence (i.e., machine learning, deep learning, and reinforcement learning) in MEC-assisted video streaming services.

\end{abstract}
\IEEEpeerreviewmaketitle

\section{Introduction}
\label{sec:intro}


The emergence of 5G brings substantial improvements in quality of service by achieving higher throughput and lower latency. These advantages enable network providers to tailor the quality of experiences for new use cases across different vertical markets. However, these new capabilities on one side are promising to transform industries, but on the other side, the massive wave of new devices and bandwidth intensive multimedia applications supported by 5G, trigger new challenges. In the context of video streaming, the improved cellular bandwidth has enabled novel use cases of video streaming e.g. ultra HD (e.g. 4K and 8K) videos, 3D videos, $360^{\circ}$ videos, virtual and augmented reality and interactive video streaming. These new classes of video streaming \khan{put} forward new demands in terms of reliable computation and reduced delay to match the desired quality of experience. These services traditionally rely on cloud computing in which the data storage and computational resources reside at the cloud servers. However, as the network scales, the increasing number of users requests can lead to significant increase in latency. Several factors add to the end-to-end latency i.e., delay caused by offloading the computation to the cloud, delay of processing and queuing at the cloud server and network communication delay to transmit the requested video back to the user. Furthermore, as the number of requests grows, meeting the delay requirement of \asif{latency-sensitive} applications becomes challenging.
\par
While latency is the key factor in the context of video streaming, there are a number of other benefits that can be achieved using MEC as compared to the traditional cloud computing paradigm. For instance, popular video streaming services such as Netflix \cite{netflix}, Hulu \cite{hulu}, and Amazon Prime \cite{amazon} create a heavy load on network. MEC can be used to cache popular content closer to the end users for a smoother experience. In cloud-based video streaming, the content is stored relatively far from the user geographical locations leading to a poor user's experience when there is a congestion along the end to end route. Cloud-based streaming services also suffers from quality of service (QoS) degradation when the users requesting similar video content grows abruptly. \par

To alleviate the issues of traditional cloud service models, mobile edge computing is emerging as a promising solution. The primary benefit of edge computing is to bring the storage and computing capabilities closer to the end users i.e., at the edge of the network. In MEC assisted 5G networks, the storage and computing services can be deployed at the network edge (i.e., within the RAN) to enable network providers better handle \asif{latency-sensitive} services. 
\asif{
MEC can bring a range of benefits to users as well as network providers. By providing storage and computing resources at the edge of the network (i.e. RAN), communication latency is reduced. Content providers can use edge storage capability to cache the content locally which reduces the bandwidth resources on the backhaul links. Distributed computing and storage resources improve resiliency and availability using collaboration among edge servers. Users can access content at lower costs available locally. Edge computing can also enhance security and greater improve scalability due to the fact that any attack on an edge server will affect the users connected to that server and not the whole network. These benefits are of MEC are further detailed in Section \ref{sec:mec_overview}.
}

MEC is still an emerging area and a huge amount of research work on mobile edge computing is in progress. MEC can help revolutionize several sectors including but not limited to, enterprises, smart buildings, healthcare, vehicles-to-infrastructure (V2I) services, Internet of Things, video streaming and virtual and augmented reality. Several survey papers exist to summarize the ongoing research efforts on mobile edge computing. A large number of these surveys provide a high-level broader view of the research area covering different aspects e.g. architectural development, functional overview, data caching, computational offloading, potential use cases, opportunities and challenges \cite{cao_2020, lin_2020, shakarami_2020, de_2019, khan_2019, hassan_2019, abbas_2018, wang_2017, baktir_2017, ahmed_2017, taleb_2017, mao_2017, ahmed_2016, shi_2016}. Some of these works cover specific aspects e.g. computation offloading \cite{lin_2020, prerna_2020, jiang_2019, lin_2019}, communication \cite{mao_2017_2} and security \cite{xiao_2019, yang_2019, liu_2019, zhang_2018, roman_2018, shirazi_2017}. There are few studies covering edge computing research in particular applications such as IoT \cite{marjanovic_2018, porambage_2018,yu_2017}, whereas some of these aim to cover the state-of-the-art scientific contributions of integrating edge computing with sophisticated machine learning and deep learning \cite{wang_2020, deng_2020, shakarami_2020, carvalho_2020, chen_2019}. A summary of these surveys is presented in Table \ref{tab:surveys}.

\begin{table*}[!h]
    \centering
    \renewcommand{\arraystretch}{1.2}
    \asif{
    \caption{Comparative study of this paper with existing surveys on edge computing. The symbol \checkmark\; indicates a publication is in the scope of a domain; \xmark\; marks papers that do not directly cover that area, but from which readers may retrieve some related insights; $\square$\; indicates the topic is covered but in less depth.}
    \label{tab:surveys}
    \colorbox{myyellow!0}{
    \input{tables/surveys2}

    }
    }
\end{table*}

Although most of the aforementioned survey papers (summarised in Table \ref{tab:surveys}) provide a detailed review of research efforts on mobile edge computing, we realise it is worthy that the huge amount of research works on MEC-assisted video streaming services shall be reviewed and summarized. 
\khan{In this survey, we provide a comprehensive review of the state-of-the-art in mobile edge computing for video streaming use cases. The paper describes different types of video streaming services and the associated challenges, and explains how these applications can benefit from using edge computing. Although we aim at focusing on video streaming, nevertheless we included reference works which are not explicitly proposed for video streaming use cases, but these have good correlation and are readily applicable in the relevant use-cases. We dedicate separate sections for novel research contributions in the area of cooperative device-to-device (D2D) communication and machine learning and their respective benefits and applications in MEC-assisted video streaming.}
\par
\begin{figure}[htbp]
    \centering
    \includegraphics[width=0.9\columnwidth]{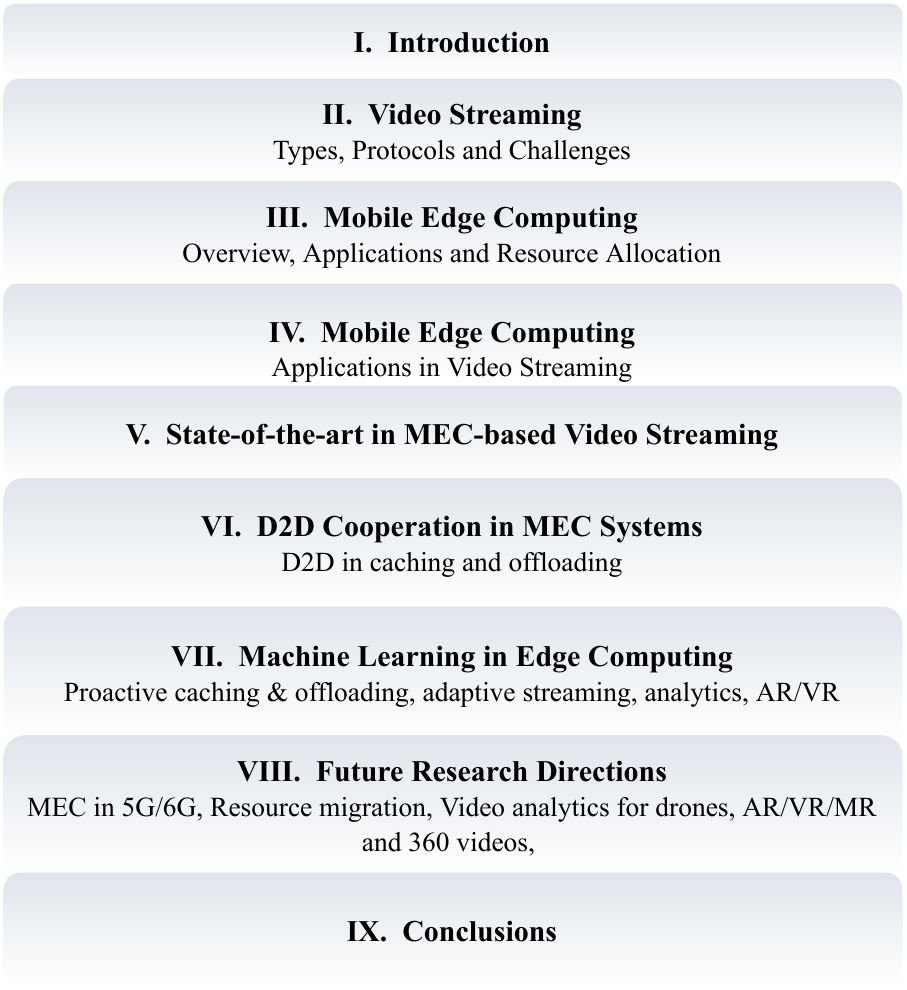}
    \khan{
    \caption{Paper Organization: Illustration of various topics covered in each section.}
    }
    \label{fig:paper_org}
\end{figure}

\khan{
The organization of the paper is illustrated in Figure \ref{fig:paper_org}. Section \ref{sec:video_stream} presents an overview of video streaming services outlining different types of video streaming services, listing streaming protocols and video streaming challenges in terms of the quality of experience. Section \ref{sec:mec_overview} provides a brief overview of mobile edge computing, explaining the architecture, functional overview, fundamental definitions and related concepts. Section \ref{sec:mec_stream} explains how MEC can improve video streaming services in different streaming use cases. A detailed review of the state-of-the-art on mobile edge computing for video streaming is presented in Section \ref{sec:mec_soa}, with separate subsections dedicated for research contributions in edge caching and computational offloading. Sections \ref{sec:mec_d2d} and \ref{sec:mec_ai} summarise the most recent developments on edge computing techniques using cooperative networking and machine learning respectively. The abbreviations and acronyms used in this paper are listed in Table \ref{tab:acronyms}.}

\begin{table}[htp]\centering
\renewcommand{\arraystretch}{1}
\khan{
\caption{Summary of Acronyms.}\label{tab:acronyms}
}
\colorbox{myyellow!0}{
\input{tables/abbrev}
}
\end{table}

\section{Video Streaming - Types, Protocols and Challenges}
\label{sec:video_stream}
Video streaming is the method of viewing videos without downloading the media files.

\subsection{Types of Video Streaming}
\label{subsec:video_types}
Video streaming is not only at the forefront of entertainment industry but also is transforming several other sectors such as enterprises, education, retail, tourism, transportation and healthcare. The unprecedented use cases of video streaming are reshaping the Internet, while at the same time, network and content providers are doing huge investments in improving user experience in video streaming services. Generally, video streaming applications can be broadly categorized as Video on Demand (VoD) and Live video streaming. 

\subsubsection{VoD Streaming}
VoD streaming allows users to watch stored videos from any Internet-connected devices at any suitable time. In VoD streaming, content can be prefetched, stored and edited before it is distributed.
\khan{Popular applications of VoD streaming are Netflix \cite{netflix}, Apple iTunes Store \cite{itunes}, and YouTube.}
There are three popular models of VoD streaming in today's video streaming world.

\subsubsection{Live Streaming}
\begin{figure*}[h!]
    \centering
    \includegraphics[width=18cm]{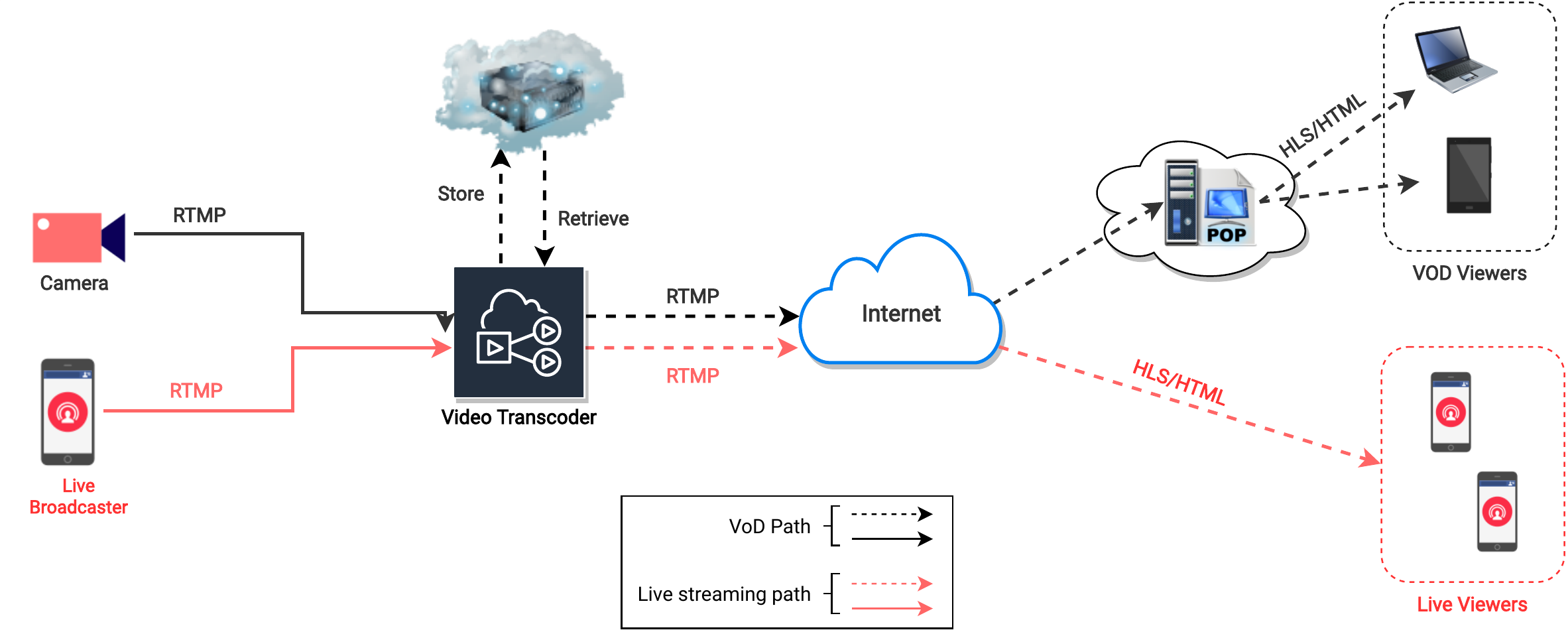}
    \caption{Video on Demand (VoD) versus Live streaming.}
    \label{fig:stream_types}
\end{figure*}

Unlike VoD streaming, in live video streaming the video is distributed in real-time directly from the origin device to the destination, without first storing it on a server. Live videos are more sensitive to network delays as compared to VoD streaming. The use of live video streaming is increasing over time \cite{cisco_air} due to multiple reasons such as the availability of high quality cameras in the modern smartphones that enable users to shoot high quality videos, and the high data rates of 5G, enabling the users to share videos over the internet in real time. Live video streaming has many attractive applications such as E-Sports and Game Streaming \cite{league_of_legends, counter_strike, overwatch, dota2, call_of_duty}, virtual reality and augmented reality, user-generated live streaming (e.g. Periscope \cite{periscope}, Youtube Live \cite{youtube_live}, Twitch \cite{youtube_live_twitch}, Facebook Live \cite{facebook_live}, Instagram Live \cite{instgram_live}, Twitter Live \cite{twitter_live} and Ustream \cite{ustream}) and online learning (e.g. Dacast \cite{dacast}, IBM Cloud Video \cite{ustream}, Kaltura \cite{kaltura}, Vimeo Live \cite{vimeo_live} and  Panopto \cite{panopto}).
\eb{
In the following, we categorize the live streaming by application scenarios:
\begin{itemize}
\item Conventional live streaming videos: This type of applications, including Twitch, YouTube live, Facebook live, allows random users to capture live videos using their handheld devices. In such applications, the live system needs to handle a huge volume of videos from broadcasters to thousands of viewers located all over the world in a very short end-to-end latency (e.g., 100 ms \cite{rev1}). Moreover, to provide the required formats to viewers based on their preferences, the capability of their devices, and the quality of the network, the videos should be transcoded. This transcoding task needs to be done in real-time and adapts on the fly to the available computing resources to match the spontaneous broadcasts.
\item Streaming 360$^{o}$ (or panoramic) videos: This type of streaming is more challenging than conventional videos as it requires higher bandwidth availability due to the volume of the content. Moreover, such live stream allows a flexible interaction, i.e., the user can potentially move and expects to see different views of the panorama. Therefore, the latency and delay variance are even stricter in order to update the display without any motion sickness. This delay is called motion-to-photon latency and should not exceed few milliseconds, in order to provide a smooth experience \cite{rev2}.
\item VR/AR live streaming: The Virtual Reality (VR) is a simulated experience that can be fictional or similar to the real world. Multiple features characterize the VR systems, which are the imagination, interaction, and immersion, in addition to the integration of the VR user in the virtual environment. The VR experience requires a device with a screen, computing components and sensors that track the user’s motions. The VR live streaming has the same requirements as conventional and 360$^{o}$ live videos, including those related to network parameters (e.g., bandwidth, latency, stalling,  buffering time, and bitrate switching) and the application parameters (e.g., video quality, frame rate, and resolution) with higher constraints on the delay variance and image freezing to avoid motion sickness. Besides these requirements, the VR system depends also on physical environment factors (e.g., sounds, objects locations, and lighting) and user’s profile (e.g., gender and length), which highly influence the quality of the experience \cite{rev3}. The Augmented Reality (AR) is also a real-world interactive experience, where real or fictive objects are enhanced and integrated into the user’s experience. In terms of system requirements, combining the real world environment with the augmented objects is very critical, as the AR algorithm has to superimpose these objects into the coordinates of the user’s location with a high accuracy. 
\\
Even though the AR and VR technologies brought new opportunities in various industries and applications (e.g., medicine, maintenance, sports, teaching, tourism and architecture.), they also added new challenges related to image display, delivery and content storage.
\end{itemize}
}
\asif{
Figure \ref{fig:stream_types} illustrates a generic functional architecture of VoD streaming and live streaming. In VoD streaming (black lines), the videos are recorded first, then transcoded to one or more bit rates versions, and then stored on a cloud server. The viewers can request and access these stored videos anytime using different protocols such as HLS. In most of the video deployments, there are also geographically distributed servers (also known as point of presence (POP)) that cache the videos from the remote cloud and serve them to the user when requested. In the case of live streaming (red lines), videos are transmitted to the end user without storing them first on the cloud. However, videos are still being transcoded in live video streaming, as different users can request different bit rate versions of the video based on the quality of the wireless channel.
}
\par

Both VoD and Live video streaming are likely to co-exist as both have different applications and use cases. VoD gives users the opportunity to watch videos anytime, anywhere and from any device using the Internet. On the other hand, live streaming provides great convenience to users to share videos in real time without first recording them. 


\subsection{Video Streaming Protocols}
Video streaming protocols can be broadly categorized as (1) Push-based (non-HTTP) and (2) Pull-based (HTTP-based) protocols. Push-based protocols are traditional streaming protocols in which the server and client first establish a connection before transmitting data. These include RTMP (Real-Time Messaging Protocol)\cite{rtmp} and RTSP (Real-Time Streaming Protocol) \cite{rtsp}, and SIP (Session Initiation Protocol). 

Since 2010, pull-based protocols have been introduced e.g. HLS (HTTP Live Streaming) \cite{hls}, Low Latency HLS \cite{llhls}, MPEG-DASH (Moving Picture Expert Group -Dynamic Adaptive Streaming over HTTP) \cite{dash}, CMAF (Common Media Application Format) for DASH \cite{cmaf}, MSS (Microsoft Smooth Streaming) \cite{mss}, Adobe HDS (HTTP Dynamic Streaming) \cite{hds}, SRT (Secure Reliable Transport) \cite{srt} and WebRTC (Web Real-Time Communications) \cite{webrtc}.

\begin{figure}[htbp]
    \centering
    \includegraphics[width=0.9\columnwidth]{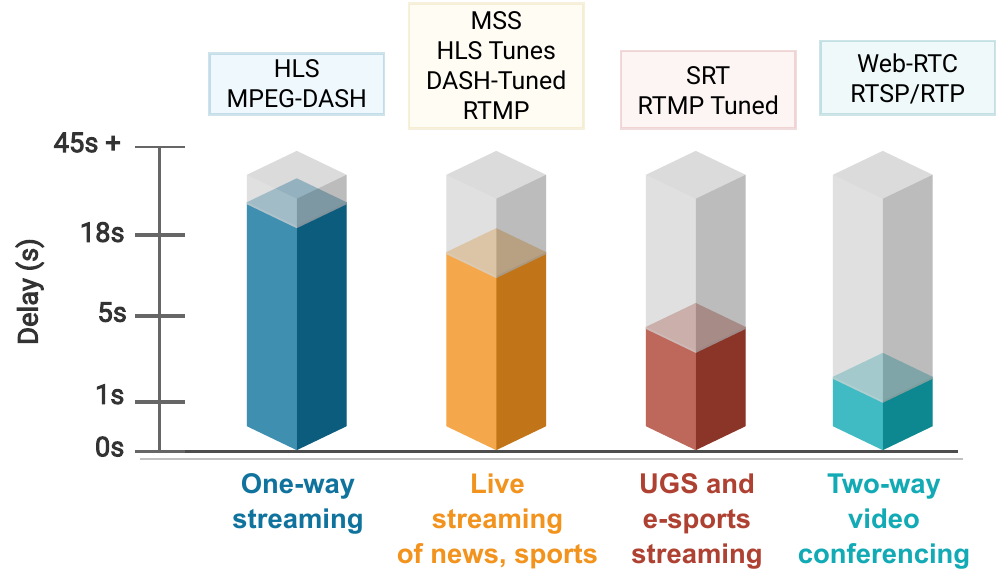}
    \caption{Video Streaming Protocols - Delay Comparison.}
    \label{fig:stram_delay}
\end{figure}

HLS is the most widely used pull-based streaming protocol \cite{stream_survey1} supported by many media players, web browsers, devices, and streaming media servers. Low-Latency HLS \cite{llhls} is an improved version of HLS which provide low latency (up to 2 seconds). The Adobe HDS was the first adaptive bitrate (ABR) protocol. Subsequently, Microsoft developed its own adaptive bitrate protocol for video streaming i.e., MSS in 2008 to deliver on-demand video of the 2008 Summer Olympics. MPEG-DASH is an open standard adaptive bitrate streaming protocol, whereas CMAF was developed by joint collaboration of Microsoft, Apple and MPEG to simplify the streaming services. The most recent addition to this list is the open standard WebRTC framework. Today HLS is the standard platform for Apple whereas MPEG-DASH is the international standard \cite{dash} for video streaming. Table \ref{tab:stream_proto} presents a summary of various video streaming protocols. A comparison of streaming delay (i.e. media transfer delay) of these protocols is illustrated in Figure \ref{fig:stram_delay}.

\begin{table}[htbp]
    \centering
    \caption{Video Streaming Protocols.}\label{tab:stream_proto}
    \colorbox{myyellow!0}{

\input{tables/stream_proto}
    }
\end{table}

Video streaming protocols can be generally categorized (based on latency) as (i) High Latency ($>$ 18 seconds), (ii) reduced latency (10-18 seconds), (iii) low latency (4-10 seconds) and (iv) Ultra low latency (1-4 seconds) \cite{amazon_latency}.


\subsection{Video Streaming Challenges}
Video streaming applications are gaining popularity due to the inherited convenience for sharing content. However, there are several challenges in live video streaming \commentout{to provide enhanced QoE to the end user}. The three major challenges associated with video streaming services are:

\begin{itemize}
    \item \textit{Limited Bandwidth:} Modern 5G networks offer higher data rates over cellular links but the growing use of video content with a much higher resolution than before is making the bandwidth limitations more prevalent. 
    \asif{In particular, the modern 4K ($3840 \times 2160$), 8K ($7680 \times 4320$), and $360^{\circ}$ videos are bandwidth intensive \cite{itu_bt2020}. On average, the 4K videos requires bit rate of 20-50 Mbps, whereas 8K videos requires 50-200 Mbps \cite{itu_f743}. To understand the bandwidth requirement of these videos, a 60-frame 4K video consumes 1GB to 10GB of traffic per minute, and a 20 min 4K video requires almost 100 GB of traffic \cite{itu_f743}. The use of these high resolution videos is also continuously increasing.
    \khan{It is estimated that by 2023, 66\% of connected flat-panel TV sets will be 4K. \cite{cisco_vni}}
    The $360^{\circ}$ videos are even more bandwidth intensive as in these, pixels are transmitted to users from every direction. The efficient transmission of such huge volumes of bandwidth intensive videos is a challenge yet to continue. \khan{As also indicated in \cite{Saad2020}, 5G systems fall short of providing a full immersive extended reality (XR) experience due to lack of support for ultra-low latency and higher data rates in such applications.}
    }
    
    \item \textit{Latency:} Video streaming in general and live streaming in particular are sensitive to latency. Higher latency leads to poor user experience in many applications, \khan{such as real-time online video gaming, virtual reality, and live streaming of high resolution videos such as 4K and 8K. With the existing content distribution architecture that relies on cloud computing, latency guarantees is a critical challenge \cite{mao_2017_2}. Similarly, in the next generation Industrial IoT (IIoT) and applications such as autonomous vehicles, augmented reality and medical imaging for remote surgeries, 5G cannot meet the sub-millisecond latency \cite{Yang2019b}.}
    \par
    
    \asif{
    In VoD streaming, videos are stored in remote cloud servers. The larger propagation distance between the cloud servers and the end users encounter delay deteriorate the user experience. To cope with this, content delivery network (CDN) servers are usually deployed to cache content from remote cloud servers to relatively closer servers located in different geographical areas. However, the CDN servers are deployed in relatively few locations and often not located in densely populated areas where viewers may reside. The transmission of video streams from these centralized cloud servers to viewers that are far apart, requires extensive bandwidth and can often create bottlenecks as the number of viewers grows. This results in inconsistent and higher startup delays, video quality degradation, and sometimes the inability to join popular live streams. }\par
    \asif{
    In contrast to VoD streaming live video streaming is more sensitive to network delays due to the fact that live videos are not cached and are directly transmitted from origin to the viewer.
    \khan{Live videos hold users’ attention $10-20\times$ longer than VoD videos \cite{live_stat1}.
    }
    In Facebook, live videos appear at the top of news feed. Also, users in a page receive notifications about a new live video post, thus the number of viewers increases rapidly. Furthermore, live streaming has random viewing patterns with very high peaks. The high number of simultaneous user requests can cause a problem known as “Thundering Herd problem” \cite{thundering_herd_problem}. A fine example of such sudden increase in demand of live video is the popular online trivia craze \cite{hq_trivia}, where the demand for online streams grew from zero to over one million viewers in just a matter of minutes. Another example is a 45 minutes video of two people exploding a watermelon with rubber bands, which reached a peak of over 800,000 simultaneous viewers \cite{watermelon}. Thus, the rapidly increasing viewing patterns of live videos can cause network congestion that results in higher end to end delay.}
    \asif{
    \item \textit{Jitter:} Jitter or delay variation is the undesired deviation from true periodicity of an assumed periodic signal. \khan{Jitters can be caused by fluctuations in queuing and scheduling delays \cite{Greengrass2009}}. Packets transmitted on the networks encounter different delays due to two reasons: First, packets route through the network independently, Second, network devices receives packets in queue and thus encounter different queuing delay. Thus, packets transmitted even at almost same time (consecutive packets) experience large variations in end to end delay. Jitter is a considerable issue in video streaming and can degrade QoS \khan{\cite{Bertino2003, Venkataraman2011}. Jitter requirements vary as per application ranging from $10 ms–50ms$ e.g., VoD streaming ($\leq 50ms$) \cite{itu_g1080}, videoconferencing and interactive video streaming ($\leq 30ms$) \cite{cisco_Qos}}.
    }
    
\end{itemize}

\section{Mobile Edge Computing}
\label{sec:mec_overview}
\khan{In this section, we elaborate the concept of mobile edge computing (MEC) and its key advantages to improve video streaming. First, a brief overview of MEC architecture and its benefits to the network operators and end users are presented. Then, we discuss some relevant works to understand the state-of-the-art in MEC for novel emerging services (not specifically video streaming). The state-of-the-art in MEC for novel video streaming services is detailed in Section \ref{sec:mec_soa}.
}
\subsection{Overview and Definitions}
The term Mobile Edge Computing (MEC) was first introduced by the European Telecommunications Standards Institute (ETSI) in 2014 as, "it provides information technology (IT) and cloud computing capabilities at the edge of the mobile network, within the Radio Access Network (RAN) in close proximity to mobile subscribers" \cite{etsi_2014}. MEC aims to deploy storage and computational services closer to the end users. It also enables third party applications and services at the edge of mobile networks. Recently, ETSI renamed its associated Industry Specification Group (ISG) as Multi-access Edge Computing (MEC).

\begin{figure*}[htbp]
    \centering
    \includegraphics[width=0.8\textwidth]{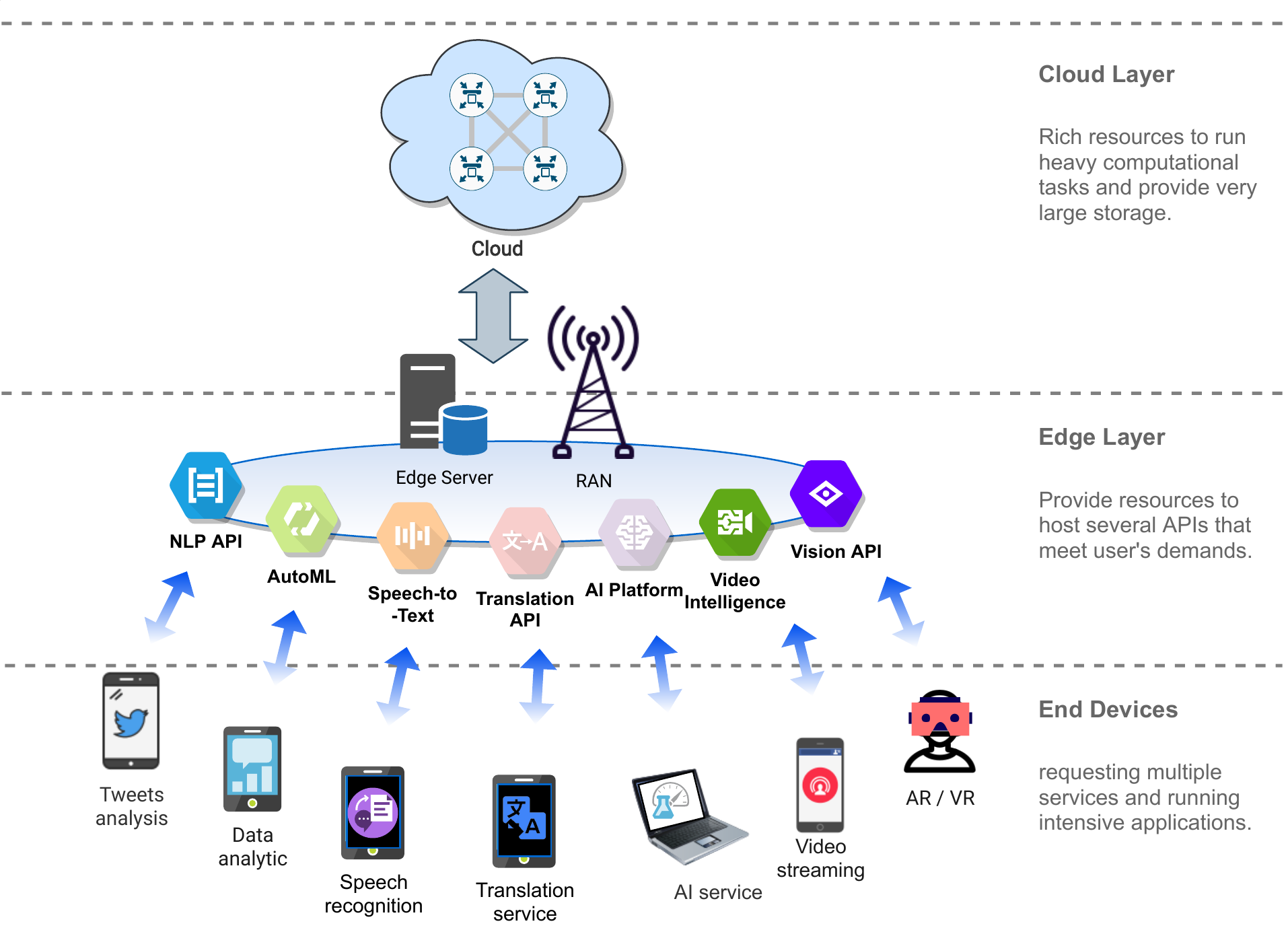}
    \caption{\asif{Mobile Edge Computing Architecture: Storage and computational servers deployed at the RAN that enables a range of services to the network users.}}
    \label{fig:mec_arch}
\end{figure*}

\asif{
Other closely related concept to MEC include cloudlets \cite{cisco2015fog, bilal_2018} and fog computing \cite{wang2020deep}. However, typically cloudlet is referred to as the architecture in which the computational servers are located closed to the user premise (not at the RAN). In fog computing, the computational capability is integrated inside the IoT gateway which connects IoT devices. The concept of edge, cloudlet and fog computing are overlapping and these terms are frequently used interchangeably \cite{mao_2017_2}. However, in the context of this paper, we will be referring to the MEC architecture in which the edge servers are located at the RAN, unless stated otherwise. Figure \ref{fig:mec_arch} illustrates the architecture of mobile edge computing. MEC allows storage and computational capabilities at the RAN level to provide several enhanced services to the end users.}

\par

MEC offers several potential advantages in live video streaming services.

\begin{itemize}
    \item \textit{Efficiency:} Using computational offloading, user devices can offload high computational tasks to edge servers for remote processing.
    \item \textit{Ultra low latency:} By storing content closer to the users, delay associated with fetching content can be significantly reduced. 
    \item \textit{Available computation:} MEC can be used to augment capabilities of other devices, thus reducing the cost of transport.
\end{itemize}

\begin{figure*}[t]
    \centering
    \includegraphics[width=0.95\textwidth]{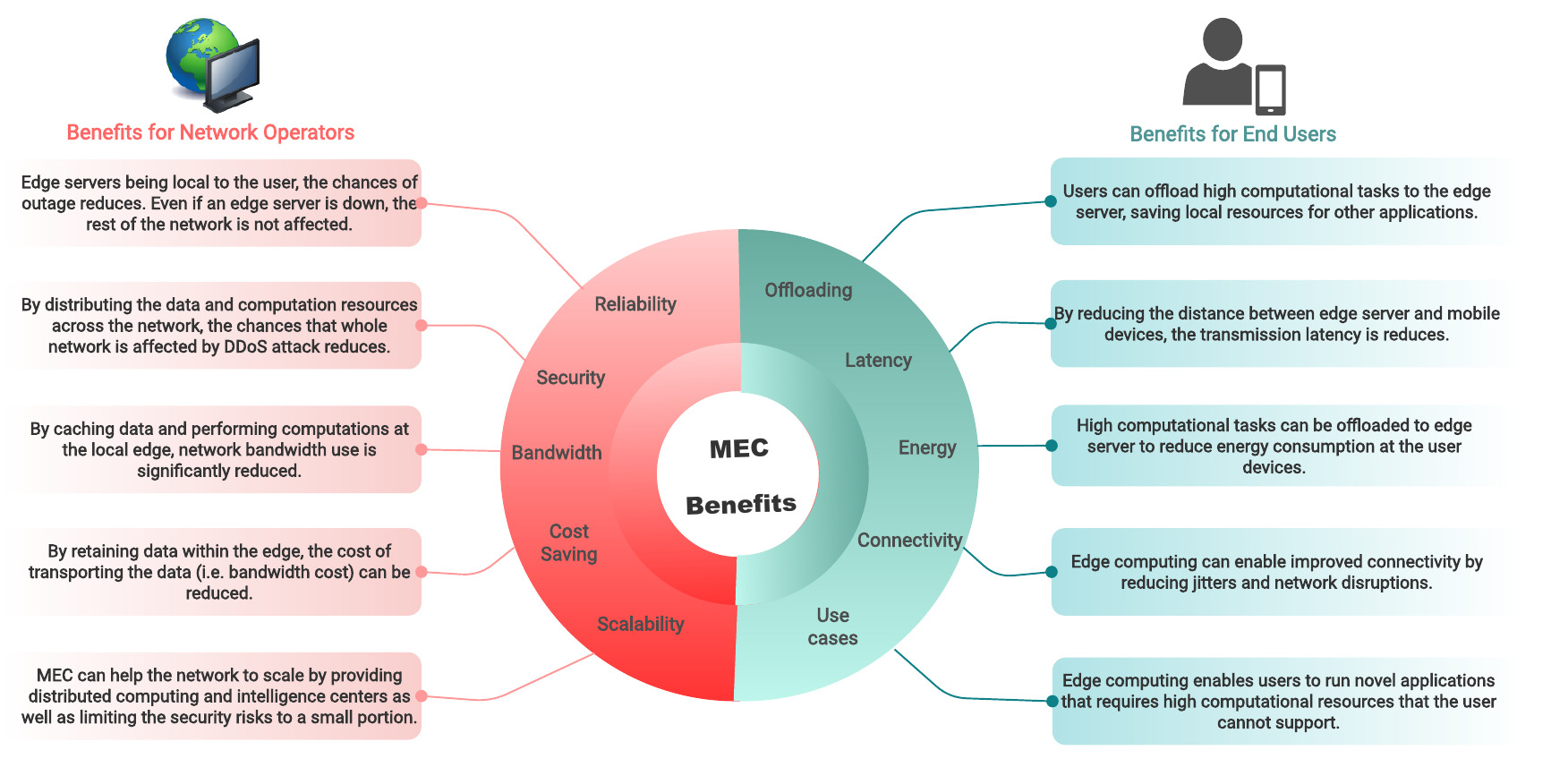}
    \caption{Benefits of Mobile Edge Computing for network operators (right side) and end users (left side).}
    \label{fig:mec_benefits}
\end{figure*}

\asif{ 
Figure \ref{fig:mec_benefits} illustrates the various benefits of MEC for end users as well as network operators in details. As depicted in the figure, these benefits are generally categorized into two categories: benefits to network providers and benefits to end users. Network providers can improve services reliability. As edge servers are distributed at RAN level, thus if one edge server is down or congested, users connected to other edge servers are not affected. This is in contrast to cloud computing which has computing resources at one or fewer locations. Similarly, denial of service (DOS) attacks on MEC servers can affect users only the users at the effected servers, making edge-based applications more robust to DOS attacks. MEC also helps to alleviate network congestion at the backhaul links, by caching and processing locally at the edge servers whereas, in MCC the data has to be fetched from remote cloud servers. When the data is stored locally, more economical content can be provided to the users, providing more business opportunities to network and content providers. As compared to cloud-based centralized services, edge-based services are more reliable and secure, hence scalability comes as an inherited benefit provided by MEC. 
}
\par

\asif{
In addition to the aforementioned benefits to network providers, MEC also bring a range of benefits to the end users. Users can offload their computation-intensive tasks to the edge server. By offloading computation and fetching locally cached content, the end-to-end latency experienced by end users can be significantly reduced. As mobile users are battery powered devices, mobile users can also exploit edge-based processing to save their energy consumption. When content is cached locally at the RAN (i.e. available at lower propagation distance or even at single-hop), video packets can be delivered with minimum delay and relatively less variations in packet delays, thus improving connectivity and reducing jitter. With the power of edge computing, mobile users can be enabled to run novel applications such as computationally intensive AI applications.
}

Mobile Edge Computing offers two major services i.e., providing (i) storage facility and (ii) computation resources, in the user proximity primarily to reduce latency with a range of associated benefits as illustrated in Figure \ref{fig:mec_benefits}. Using MEC, users can access locally stored content with low latency, and offload their computational tasks to the edge server for fast processing. In the following, the two concepts are explained in details. Furthermore, a brief review of recent research works in these areas is presented. (Note: The review of the state-of-the-art of MEC in video streaming services is presented in Section \ref{sec:mec_soa}). 

\subsection{Edge-based Caching}
Traditional cloud-based architectures store millions of videos in relatively large-sized servers in few geographical locations. Edge caching is the process of fetching content from these remote cloud servers and storing at the local edge servers. Edge caching reduces data traffic transported over backhaul links and reduces the content delivery time. Edge caching techniques can be categorized in different ways e.g. proactive versus reactive caching, independent versus collaborative caching.

Proactive caching refers to fetching content before arrival of the request usually based on the video popularity or probability of the request, whereas reactive caching is post-request content fetching. Independent caching refers to the capacity of an edge server to solely decide content fetching based on local or central information, whereas in collaborative caching (also referred to as coordinated caching), multiple edge servers can provide their cached content to each other.

Any caching algorithm has two parts, i.e., \textit{content fetching} and \textit{cache replacement}. Content fetching refers to the process of bringing content from the remote server (origin or cloud storage) and storing it into the local/edge server. Cache replacement refers to the process of selecting content that need to be removed from the local/edge server if there is limited storage size available to store the newly fetched content. Content caching in traditional networks such as content centric networks have been extensively investigated in several studies \cite{khreishah_2014, khreishah_2015}. 
In edge computing, the caching problem is typically studied as a joint problem with the offloading decision \cite{tran_2017,Ge2017}. A detailed review of caching schemes in video streaming context is provided in Section \ref{sec:mec_stream}.

\subsection{Edge-based Computation}
\begin{figure*}[htbp]
\centering
\includegraphics[width=0.9\textwidth]{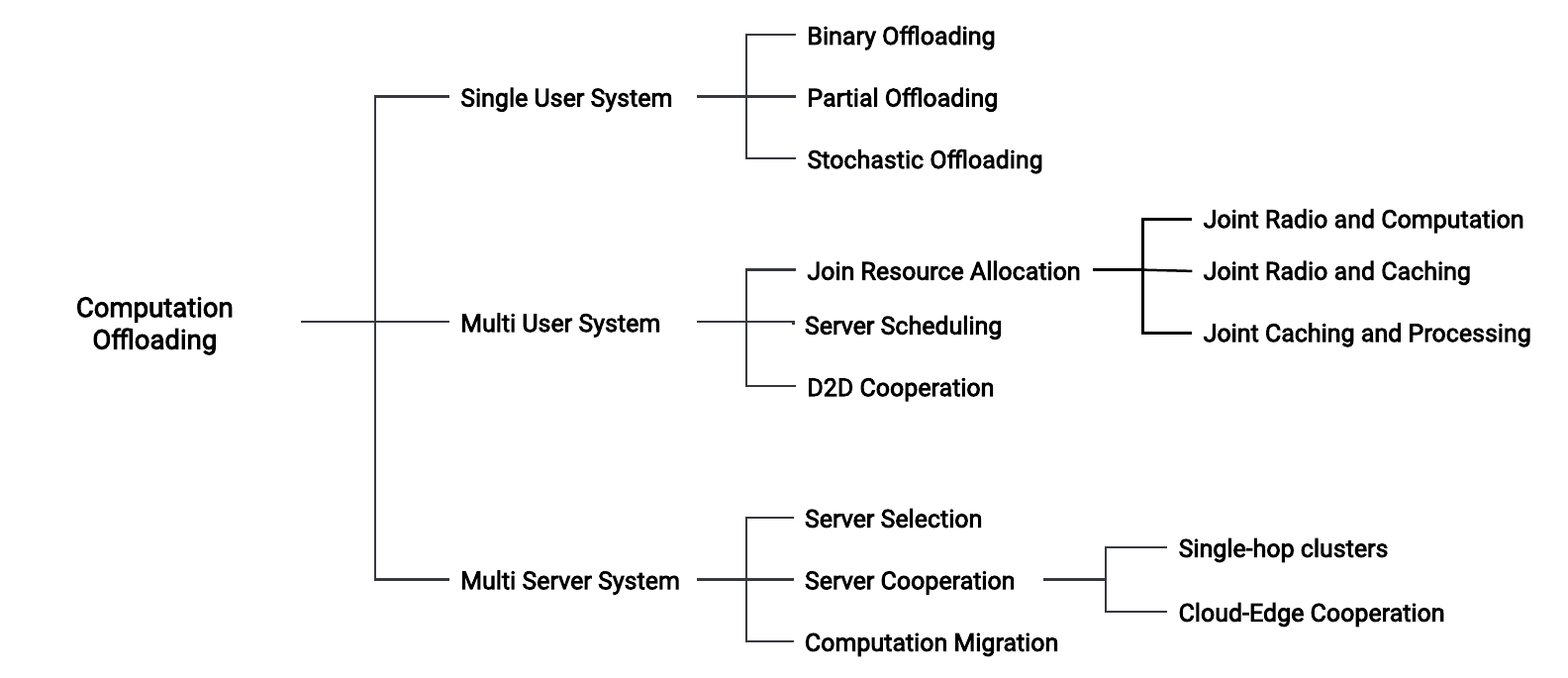}
\caption{Computational Offloading in Mobile Edge Computing.}
\label{fig:resource_alloc}
\end{figure*}

Computation at edge nodes can be implemented in several ways depending on the design objective. Common design objectives for implementing edge-based computation and optimization include energy consumption, computation latency or bandwidth utilization. In this section, we present a brief overview of different computation offloading models in mobile edge computing and a summary of some representative works of each category. As illustrated in Figure \ref{fig:resource_alloc}, the computation models are categorized into three classes. \khan{A similar classification of computation offloading in MEC networks can be found in \cite{mao_2017_2}}.

\subsubsection{Single User Systems}
This is the simplest computation offloading model that consists of a single edge server and a single user. Although real networks always have multiple users, the purpose of the model is to simplify the computation offloading decision by considering the task of every single user independently. The model decides whether a particular task of a user should be computed locally at the device or offloaded to the edge server. The model can be sub-classified as (i) \textit{binary offloading} and (ii) \textit{partial offloading}. 
\par
If the whole task is computed locally or offloaded wholly to the edge, this is referred as "binary offloading". In contrast, when a user device computes a part of the task locally whereas offload the remaining part to the edge, the model is referred as partial offloading.  Examples of binary offloading include \cite{barbarossa_2014, zhang_2013, Hao2019}.
In \cite{barbarossa_2014}, authors investigate the simple case of single user to analyze computation offloading while minimizing the energy consumption at the mobile device under a computational rate constraint. The model considers various parameters including device's and server computational power and the communication bandwidth to take the offloading decision. A task is offloaded to server if the energy consumption by the user device for offloading the task is less than the energy needed for local computation. \khan{The energy consumption is studied as a function of distance between the device and edge server. The results show improved energy performance at shorter distance which increases when the distance increases. The delay in this work is considered as a hard constraint.}
In \cite{zhang_2013}, authors propose an alternate method to solve the binary offloading decision \khan{to run mobile application locally or at cloud server. The objective is similar to \cite{barbarossa_2014} i.e., to minimize the device's energy consumption, while considering soft delay deadlines. The energy consumption in this work is modeled as a function of CPU frequency:}
\begin{equation}
    \khan{ \xi(f) = k f^2 }
    \label{eq:energy1}
\end{equation}
\khan{
where, $k$ is the switched capacitance (set to $10^{-11}$) depending on the processor chip architecture. To minimize the energy in local computing mode, the CPU frequency is adjusted, whereas in remote computing mode, the data transmission rate is adjusted. A limitation of this work is that in local mode, the energy consumption is reduced by reducing CPU cycle frequency which means the application will run slowly, which make it unsuitable for heavily intensive applications with delay constraint.} 
In \cite{Hao2019}, binary task offloading is proposed to maximize the revenue of the service provider. \khan{This paper does not focus on the QoS requirement of the user/application}. Other works on binary offloading can be found in \cite{kumar_2010, huaming_2013}.

Partial offloading schemes in MEC systems have been investigated in \cite{wei_2018, wang_2016, jia_2014}.
\khan{In \cite{wei_2018}, authors proposed a partial task offloading model in which a single task is divided into sub-tasks and then each sub-tasks is either computed locally or offloaded to the edge. The task offloading depends upon the local computation resource available, the communication channel capacity and the queue size in the edge server. The offloading decision is optimized using a greedy algorithm (called as Select Maximum Saved Energy First (SMSEF) algorithm) to maximize the energy saving at the mobile devices. However, this scheme does not minimize the task completion delay, but only ensures that the delay deadline is met.}
In \cite{wang_2016}, authors proposed a partial computation offloading scheme by jointly optimizing the computational speed, transmit power and the offloading ratio. The study aims at minimizing both the energy consumption and the computation time. \khan{The energy consumption is modeled similar to Eq. \ref{eq:energy1}.}
In \cite{jia_2014}, authors proposed a heuristic that makes an online offloading decision on sequential tasks. The algorithm aims to minimize the completion time of the application. \khan{The task completion time for local and remote execution is modeled as:}

\khan{
$$
\text{Execution time} =
    \begin{cases}
        \frac{s}{\mu_m} & \text{Local} \\
        \frac{s}{\mu_c} + \sum{\frac{s_i}{\mu_c}} + \frac{d_i,j}{R} & \text{Remote}
    \end{cases}
$$
}
\khan{where, $s_i$ is the task size (as number of instruction), $\mu_m$ and $\mu_c$ are the CPU capacity of mobile device and the edge server respectively, and $R$ is the data rate of the wireless channel.}
Other works on partial offloading can be found in \cite{yang_2012,huang_2012,munoz_2015, boutheina_2019}.

Binary offloading is preferred for problems involving similar tasks and in which user device has strong channel conditions to the base station for fast computation. The benefits are to reduce the overall latency reduction of the system and energy saving at the user device. On the other hand, partial offloading is a preferred model for problems involving heterogeneous tasks, thus offloading computational intensive portion of the task to the edge whereas perform the remaining tasks locally. Similarly users can decide to offload tasks of small transmission payload to the edge to reduce data loss.
\par
The aforementioned two single user MEC models are used when the task arrival process is deterministic. However, in some cases, the task arrival is random i.e., tasks arrive at the processor queue at a random rate. In such a case, the offloading decision is also dynamic and is referred to as \textit{stochastic offloading}. Stochastic offloading becomes a necessary requirement in some problems e.g. if the random task arrival rate is leading to buffer overflow, dynamic offloading can help to avoid this by randomly offloading tasks to the edge when the task arrival rate increases. Another use case is to exploit the randomness of the wireless channel to design channel-aware offloading schemes in which tasks are offloaded only when the channel supports the required communication delay. Example works on stochastic offloading can be found in \cite{yang_2018, haouari_2019_2}. \khan{As compared to deterministic task arrival rates, stochastic tasks arrival requires more robust techniques to optimize the offloading decisions. Authors in \cite{yang_2018} use Q-learning to perform the offloading such that the tasks are completed within the delay deadlines, thus achieving the end-to-end reliability. Similarly, authors in \cite{haouari_2019_2} studied the resource allocation problem in dynamic scenarios. The objective is to reduce the unnecessary resource allocation, maximize the QoE and minimize the network cost to the content provider in a multi-edge servers environment. The authors used several machine learning (ML) algorithms (Long Short Term Memory (LSTM), Gated Recurrent Unit
(GRU), Convolutional Neural Network (CNN), MultiLayer Perceptron (MLP) and XGboost) on synthetic datasets to evaluate the effectiveness of each model. The use of ML based techniques achieves better performance due to their capability to accurately estimate the stochastic parameters in real-world applications.}

\subsubsection{Multi User Systems}
Multi users system refers to the computation model in which multiple user devices share a  single edge server. In such systems, computation offloading can be implemented in various ways. For instance, the tasks offloaded by multiple users can have different priorities, hence the server must schedule the tasks computation according to the priorities of the tasks. Similarly, different users have usually different channel conditions and consequently support different data rates, hence a "joint resource allocation" model which jointly optimizes the offloading and radio resources is desired to improve system-wide performance. Examples of joint optimization of radio and offloading in multi users MEC systems are studied in \cite{barbarossa_2013, ren_2019}.
\khan{In such kind of problem, one of the parameter is taken as a hard constraint, while optimizing over the others. For instance, in \cite{barbarossa_2013} the mobile energy consumption is minimized under average latency constraint, whereas in \cite{ren_2019} the weighted-sum of latency is minimized while considering/satisfying the mobile devices' energy consumption requirements. The joint optimization over two parameters does not preclude to consider other parameters such as computational and storage resources.}

As the edge server also provides caching facility to store content, the joint optimization of caching and processing can provide several benefits such as service provider revenue or content retrieval latency and storage utilization. Joint caching and processing systems are studied in \cite{tran_2017,Ge2017}. \khan{In joint caching and processing, the objective is to find the optimal location (i.e., edge/cloud servers) for storing and processing the content. For instance in \cite{tran_2017},  authors consider jointly the storage and processing resources to minimize the backhaul network cost of serving all requests. This is done by jointly determining the cache placement and processing scheduling using Integer Linear Programming (ILP).}

In multi user systems, when users offload tasks to the edge server, the tasks may have different priorities. To meet the QoS of such applications, the edge server may schedule these tasks based on the latency requirements. Hence, latency sensitive tasks are computed first, followed by latency-tolerant tasks. Examples of server scheduling include \cite{molina_2014, guo_2016, mehrabi_2019}.
\khan{Authors in \cite{molina_2014} proposed task scheduling based on the energy saving opportunity. Thus, a device offloads the task to the server when the offloading can achieve more energy saving than local computing. However, the offloading method in this work is binary and task splitting was not investigated. In \cite{guo_2016}, authors propose the task scheduling under strict delay requirement. The work proposes to use dynamic voltage and frequency scaling (DVFS) technique for local computing whereas adjusting transmission power for remote computing (edge/cloud) to reduce energy consumption in both modes. In \cite{mehrabi_2019}, authors further investigated the scheduling problem by jointly considering server scheduling and video bitrate selection to improve the overall QoE and fairness in resources allocation.}

Unlike the cloud computing paradigm, in edge computing, the edge servers have limited computation power. Furthermore, serving heterogeneous users requests becomes more challenging when the server has limited resources. To overcome the resource limitations, recently, MEC operations have been investigated with device-to-device cooperation. D2D-assisted MEC systems have been recently proposed in \cite{Pu2016,Wu2019,Yuan2019, xing_2019}. In D2D-assisted MEC systems improve system's overall capacity in two ways. First, users can exploit D2D cooperation to offload tasks to its neighboring devices, instead of offloading to the resource limited server. Secondly, MEC servers can itself offload computation tasks to user devices.

\subsubsection{Multi Server Systems}
Multi Server systems refer to the MEC systems consisting of multiple and typically heterogeneous MEC servers. Multi server systems offer several challenges.
The first challenge in multi-server system is the selection of server for offloading computation. Different selection criteria can be considered e.g. selecting closest server, selecting least loaded server, selecting between cloud or edge server. \khan{For example, in \cite{haouari_2019}, authors proposed a scheme to select the closest server to minimize the resource allocation cost and maximize QoE. The work used the FaceBook 2018 live video dataset to predict the viewing patterns and automatically select the edge server to minimize the startup delay. The work employs ML techniques such as Multilayer-perceptron (MLP), Decision trees (DT) and Random Forest (RF). Similarly, in \cite{niu_2019}, authors propose a load balancing and task allocation scheme using Particle Swarm Optimization (PSO). The users select the edge server based on multiple criteria i.e., current server load and distance to the edge server. }

Multi-server systems can benefit from the cooperation among the servers. Servers can collaborate with each other in several ways. For example, servers in a small geographical area connected by single hop cooperate by offloading computation to each other. Similarly, edge servers can collaborate with cloud servers, to compute delay-sensitive tasks locally (at the edge server), whereas offload delay-tolerant tasks to the cloud servers. Examples of edge-cloud cooperation are \cite{zhao_2015, ren_2019, bilal_2019}. \khan{In \cite{ren_2019}, cloud-edge cooperation is used to split tasks among edge and cloud to minimize the latency. On the other hand, \cite{zhao_2015} proposes to schedule tasks such that delay-sensitive tasks are processed at the nearest resource-constraint edge server whereas delay-tolerant tasks are offloaded to the remote resource abundant cloud servers. Similarly, in \cite{bilal_2019}, authors propose the cooperation among multiple edge servers to share contents via backhaul links when requested.}

In multi server systems, a mobile user may move away from one server and get closer to another MEC server. Consequently, the network controller can optimize computation in two ways: (i) by offloading computation to the new server or (ii) performing computation on the origin server and then forwarding the computation results to the new server. Example works on computation migration are available in \cite{wang_2014, urgaonkar_2015, chen_2016}. In \cite{wang_2014}, authors used Markov Decision Problem (MDP) to formulate the computation migration in multi-server environment. The migration decision is taken considering the distance between the users to each server and using two thresholds. The work is extended in \cite{urgaonkar_2015} by jointly considering the computation scheduling and computation migration to minimize the average transmission energy and reconfiguration cost. In \cite{chen_2016}, authors proposed to compute the tasks locally or migrate computation to remote cloud server such that the total energy consumption and latency is minimized.

Table \ref{tab:resource_alloc} provides a brief summary of representative works on resource allocation schemes in edge computing. In Section \ref{sec:mec_soa}, we provided a details review of the state-of-the-art on resource management in video streaming applications. A more exhaustive list of research contributions on resources management in MEC in generic applications is provided in \cite{mao_2017} and \cite{kumar_2013}. 

\begin{table*}[htbp]
    \renewcommand{\arraystretch}{0.9}
    \centering
    \caption{Computational Offloading Schemes in Mobile Edge Computing.} \label{tab:resource_alloc}
    \colorbox{myyellow!0}{
    \input{tables/resource_alloc1}

    }
\end{table*}

\subsection{Lessons learned and challenges}
Edge computing provides storage and computation resources closer to the mobile devices i.e., at the base station. In the caching process, content is pre-fetched and stored at the edge server. As edge storage capacity is limited, older content need to be replaced when it is reaching the available storage capacity. \khan{Content caching can be implemented in non-cooperative method such as in \cite{cha2009,ahlehagh_2014} or cooperative method \cite{Borst2010,Jiang2017}. Cooperative caching methods are more popular due to their system-wide performance gains.}  
The computation power of edge server is also limited and hence efficient resource allocation strategies are required. In the simplest case (single user system), a mobile device connected to an edge server can offload \khan{computationally} intensive tasks to the edge whereas perform simpler tasks locally. Alternatively, mobile devices can also offload sub-tasks to the edge (i.e., partial offloading). In the case of multi user systems, edge server can schedule offloaded tasks according to the latency requirements of the user's applications. 
\par
\eb{
Due to the high number of decision factors and the system uncertainties, the resource allocation problem in mobile edge networks becomes highly complex. More specifically, to optimize the MEC resource utilization, multiple decisions should be considered, including the caching, computing and networking variables. The main metric related to the caching is the hit ratio, whereas the metrics related to the computing are the latency, the throughput, and the energy consumption. On the other hand, the transmission latency, the data rate and the QoE, should be taken into account while scheduling transmission decisions. Resource allocation problems are typically addressed by formulating an optimization problem with one or multiple objective functions using different system constraints. The traditional techniques to solve such problems are included mainly under the umbrella of stochastic and convex optimizations and game theory. However, these approaches are very complex and time consuming and are not adequate for online implementation. Recently, reinforcement learning gained a lot of attention owing to its ability to solve resource allocation systems, particularly, those with dynamic, large and complex problem spaces. RL-based approaches are still in their infancy and further efforts need to be conducted to examine their performance on MEC networks. 
}
\par
Further improvements can be achieved by considering the users radio channel in the joint optimization scheme. The joint optimization of edge resources can improve the resource utilization to a certain level, however  recently, D2D collaboration has been proposed in edge computing system, which enabled mobile devices to offload computations to other resource-rich helper devices. Such types of D2D-enabled MEC systems can boost the performance of the system beyond the edge capacity limits.\par


\section{Mobile Edge Computing - Applications in Video Streaming}
\label{sec:mec_stream}
\asif{
In the previous section, we discussed the detail overview of mobile edge computing, its architecture and state-of-the-art including caching and computation. However, it is important to understand how MEC can help to improve the users' video streaming experience and/or implement novel video streaming applications and services.
}
The primary benefit of mobile edge computing in the video streaming is reducing the end to end latency to enhance user quality of experience by eliminating network lags, frames dropping and buffering. However, in addition to latency reduction, MEC alleviate the network congestion by running applications and performing the processing tasks closer to the end users. As the MEC servers are deployed at the base station (i.e. RAN), thus it allows flexible and rapid deployment of new applications and services for cellular customers. Similarly, as network service providers can authorize trusted third parties, such as application developers and content providers to deploy MEC-based services. \par

\begin{figure}[!h]
\centering
\includegraphics[width=6cm]{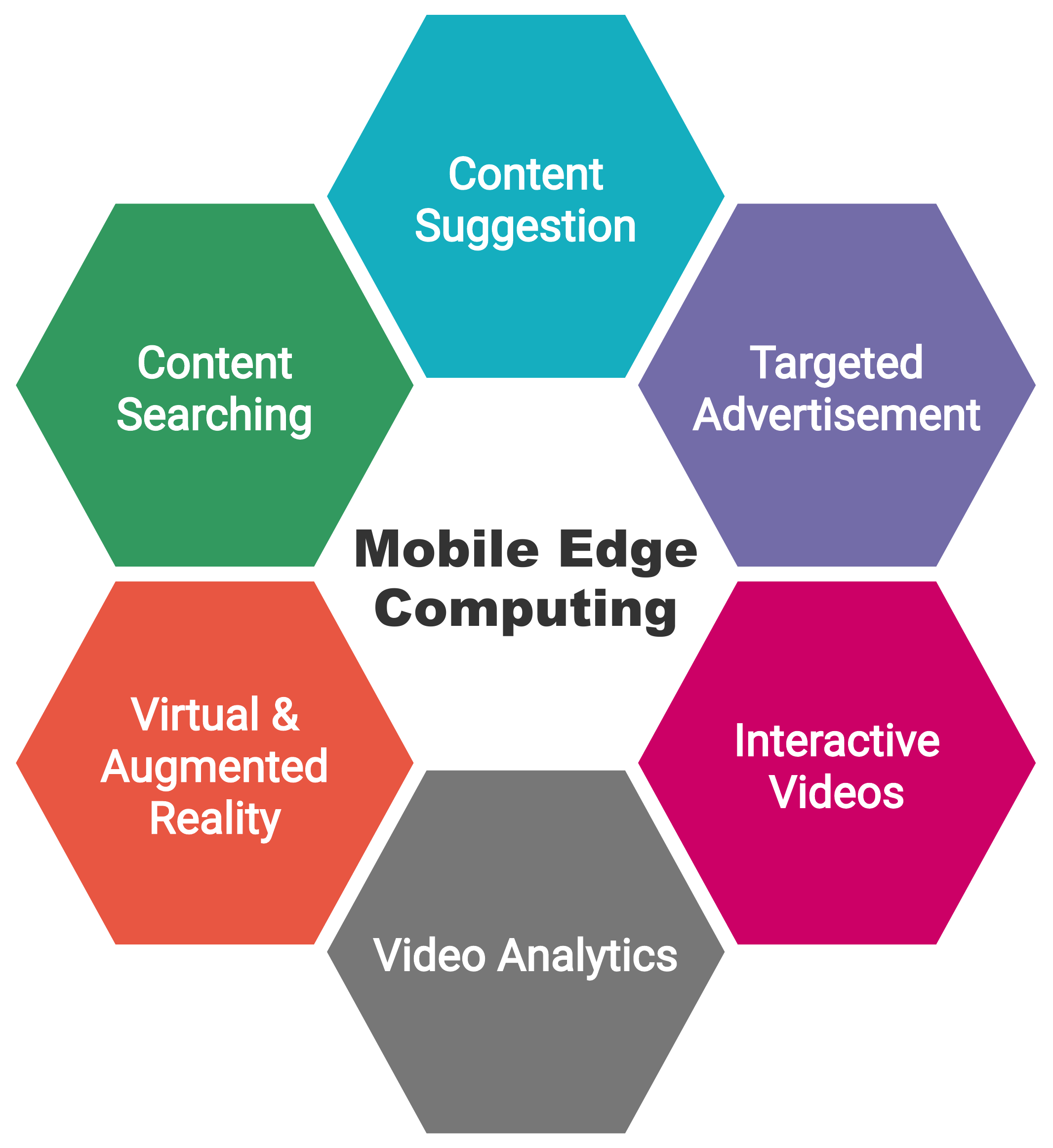}
\caption{MEC Use cases in Video Streaming Applications.}
\label{fig:mec_video_apps}
\end{figure}

\khan{This section outlines novel video streaming services in which MEC can bring potential advantages. It is evident from the previous discussion in Section \ref{sec:mec_overview} that MEC helps improve video streaming via optimal caching of contents at the network edge and provide computing resources for faster processing. However, how these inherited advantages of MEC benefit video streaming? What are the scenarios and novel services that can be realised using MEC? This section provides an overview of the attractive applications of MEC-based video streaming services. Figure \ref{fig:mec_video_apps} illustrates the most attractive applications of MEC assisted video streaming, whereas specific examples of each application category are listed in Table \ref{tab:mec_apps_stream}.
}

\begin{table}[]
    \centering
    \caption{MEC Applications in Video Streaming.}

\input{tables/mec_apps_stream}
    \label{tab:mec_apps_stream}
\end{table}

\subsection{Content Searching}
The powerful computing capabilities of the edge server can bring a more personalized search features which runs faster than traditional approaches using cloud computing. Using MEC, locally available information at the edge servers about user preferences, location, local events and incidents can help improve user search experience \cite{shi_2016, cs_mec1}. For instance, users in a specific area when search for a local event or incidents, the content searching can be optimized based on the other local users' search history and selections data saved at the local MEC server. This kind of localised content searching is only possible using MEC servers deployed in users' proximity.

\subsection{Content Suggestion} MEC offers the computational capabilities to process the data related to user preferences and activities. When such processing is done faster, the user can be served with appealing content in real time that ultimately enhancing user streaming experience \cite{cs_mec1}. The kind of location specific content suggestions is based on the search history of other local users, which is saved in the edge caches.

\subsection{Targeted Advertisement} When powerful computations are applied to user's preference data, the usually unwanted advertisements that interrupt streaming videos can be tailored to the user's interest. Advertisement agencies can benefit from the user search preferences, search history, previous purchase history, locations visited with timing information etc., to send highly targeted advertisements \cite{ta_mec1}. As an example, a user at a shopping mall would be probably buying some goods whereas another user at a hospital would be interested in a healthcare product. MEC thus allows location-based service recommendations tightly coupled with a specific place. Particularly, information of radio node to which user terminals is connected is available at MEC server that can be used to get location information of users especially when users are connected to an indoor small cell. Then an inference engine deployed at the MEC server can determine proper services and sends the related advertisements for the user at the moment.

\subsection{Interactive Video Experience} Users would always enjoy interacting with live videos e.g. displaying the statistics of a baseball game being streamed, pulling the filmography of an actor as he appears in a scene. Such user interaction always requires a level of immediacy which can be made possible using the high computation power of the edge servers available with minimum latency \cite{Bilal2017}.

\subsection{Video Analytics}\label{Analytics}
\emna{Surveillance and video analytics to detect accidents are included under the umbrella of live video streaming. More specifically, in surveillance applications, the aim is to monitor a specific area and identify potential threats within the target region. Some of the area are very critical and need 24/7 surveillance such as military borders, oil/gas off-shores, and forests exposed to potential fires. The video frames (or video streaming) are sent instantaneously to remote servers for real-time object/accident detection using AI for example. The traditional wisdom resorts to cloud or servers to compute these heavy tasks. However, video streaming destined for detection and surveillance do not tolerate high latency, such as forest fire detection that needs immediate intervention. Furthermore, used cameras are sending high-resolution video frames to cloud servers and knowing that incidents are rarely occurring, the  large  data  volume transmitted by source units has  become  problematic, particularly  for  systems  that  do not  have  stable  bandwidth  availability. Because of this tremendous amount of data, video analytics should be done at the edge of the network.} 

The huge amount of videos generated by mobile users, social media, IoT devices, scientific apparatus, satellites, and video surveillance cameras is being processed by modern computer vision techniques powered with AI. The processing can be done on-camera, which requires expensive AI-powered chips. Alternatively, to process these videos on cloud, longer delays occur as previously described (approximately 150 to 200 milliseconds). Edge computing offers a good trade-off by eliminating the need for on-camera processing to reduce cost and processing videos locally reduces delay (around 10 milliseconds). Reduced delay allows for quick detection and faster response which is required in many applications. \par
Video streaming analytic has a broad range of applications such as head counting in live streaming videos, suspicious activity detection, correlations in different video streams, combining real time information with historical context and search and rescue in live videos captured using drones. All these use cases require low latency (sometimes ultra low latency) analysis of streaming videos, thus motivates for using edge computing. Representative works in video analytics can be found in \cite{Yang2019, va_mec1, va_mec2, va_mec3, va_mec4, va_mec5, Sun2020}.

\subsection{Virtual Reality (VR) and Augmented Reality (AR)}
VR and AR are two emerging technologies that connect the physical and digital worlds. While there is a distinction between both\footnote{AR overlay digital elements such as visual content and information on your real world view, whereas VR implies a complete immersion experience, allowing users to experience things and places that actually do not exist there in the user's environment. Mixed Reality (MR) refers to combining VR and AR technologies for a richer user experience.}, both uses streaming videos to enrich user experience. \khan{There are a range of attractive use cases of both technologies such as games, entertainment, training, education and scientific areas \cite{vr_mec1}. A list of games using VR/AR technologies can be found in \cite{vr_mec2}.}\par
\emna{Recently many commercialized smart applications \cite{mirror} are proposing to create a virtual avatar of the user, that allows to try the clothes virtually.  More specifically, the user is able see his/her reflection on the screen as if he/she is looking to a mirror, where the virtual clothes are blended with the scene. This process is called blended-reality. The application users can see themselves moving, turning around, and walking while being dressed with the chosen items. The users can also see, in 360°, how the item looks like on them and if the size is appropriate. In the same context, Amazon has published a patent \cite{Amazon_patent} of its partially reflective intelligent mirror, where virtual clothes and real scenes are transmitted through the mirror to generate a blended-reality client able to see himself/herself wearing new items. This type of application, called interactive application, uses virtual reality streaming that is extremely intolerant to delay variance or image freezing, as the user wants to see his/her avatar without any motion sickness affecting the quality of experience. For this reasons, the cloud wisdom to blend the streamed video is no longer sustainable of such real-time applications and sending data to remote severs may not satisfy the latency requirements.}
\khan{MEC can be used to improve VR/AR applications for several benefits such as latency reduction \cite{vr_mec3, vr_mec4}, efficient resource utilization using device-MEC collaboration \cite{vr_mec3}, improved throughput \cite{vr_mec3}, and reducing the backhaul traffic load \cite{vr_mec5}.}

\subsection{Lessons learned and challenges}
Mobile edge computing can help improve video streaming services by providing proximal caching and computational resources for transcoding the videos. In addition, faster edge-based processing of videos, a number of benefits can be achieved such as faster content searching and personalized suggestions, targeted advertisements, real-time user interaction with videos, video analytics for surveillance, object detection and real-time assistance, and virtual reality applications such as gaming, retail, healthcare and industrial safety. \khan{It is widely believed that 5G networks fall short of meeting the unprecedented requirements of data-intensive and delay-sensitive applications such as autonomous vehicles, remote surgeries etc., thus leaving room for acceptance of edge-based solutions. The future 6G networks further mandates the use of edge intelligence in a variety of network functions and services. However, the success of MEC-based solution heavily rely upon the edge infrastructure deployment. MEC deployment will incrementally progress in a way to support as new network services are introduced. As will be explored further in the subsequent section, MEC deployment considerations include revenue as an important factor in addition to the QoS metrics (e.g., delay, storage, throughput, energy etc.).}


\section{State-of-the-art in MEC-based Video Streaming}
\label{sec:mec_soa}
\khan{In the previous section, we discussed MEC-assisted video streaming applications and services. However, to implement these services in real-world applications, providing caching and computing resources along the network edge is not sufficient. Indeed, it involves various challenges such as efficient caching strategies, optimal resource allocation, cooperation among network entities, and  tasks/requests scheduling, etc, to realise the full range of these applications. There have been a huge amount of research efforts contributed to cope with these challenges. This section focuses on covering these research works to solve the aforementioned problems. We have dedicated separate sections for the works involving device-to-device (D2D) cooperation (in Section \ref{sec:mec_d2d}) and machine learning (in Section \ref{sec:mec_ai}).
}
\subsection{Content Caching at Edge}
As discussed earlier, MEC offers proximal storage at the edge server that significantly reduces the delay to retrieve content and alleviate the congestion and bandwidth usage for network operators, resulting in overall improved performance and QoS in several services \cite{Yang2019}. However an edge server has limited storage as compared to cloud servers to store large-sized video content, hence caching schemes need to be optimized for efficient allocation of storage resources to end users. In the following, we present research contributions aiming at edge caching optimization.

\begin{figure}[!h]
    \centering
    \includegraphics[width=0.9\columnwidth]{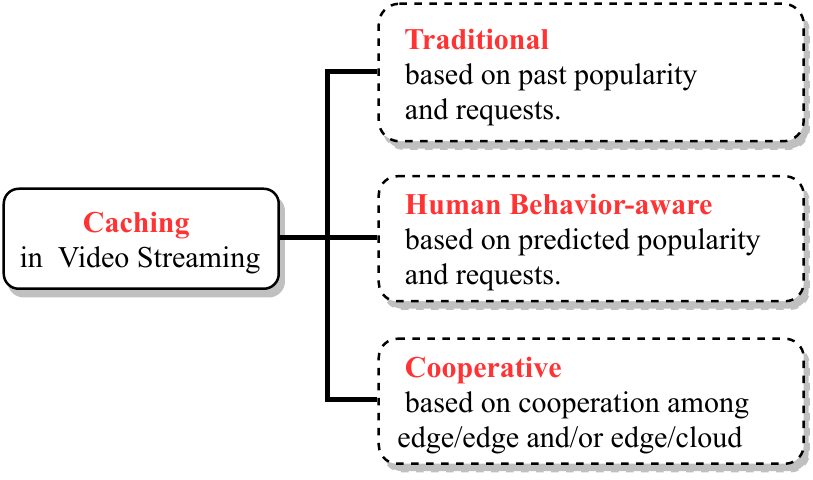}
    \caption{\khan{Caching Techniques in Video Streaming.}}
    \label{fig:caching_soa}
\end{figure}


\emna{
One of the key parameters that influences the content retrieval latency and the traffic transmitted over the network is the distance between the end-user and the remote servers. In this context, caching the video content in MEC servers is a promising solution that enhances the bandwidth efficiency and reduces the latency to serve viewers. However, the edge servers are characterized by their limited capacities in terms of storage. On the other hand,  we are witnessing an explosion of newly published video streaming. For example, in 2020, users have been uploading to Youtube around 500 hours of videos every minute \cite{youtube_stat}. Therefore, edge caching should be performed wisely. Several approaches responsible for scheduling the content caching and removal have been proposed in the literature. These approaches can be classified into two groups, namely traditional schemes and popularity/human behavior- aware strategies.
}

\emna{
\subsubsection{Traditional Caching Schemes} The traditional schemes assume that each video has a popularity assessed by the number of views and this popularity remains stable, even after passing the trend. Several policies are proposed in this context:
}

\begin{itemize}
    \item \khan{\textit{Most Popular Videos (MPV):} This technique was originally used by Hulu \cite{Krishnappa2011} for content caching. MPV is a proactive caching policy that caches the content based on the nation-wide content popularity. The shortcoming of MPV is that the cache is not updated based on the user requests or viewing patterns. MPV is suitable for large caches such as Internet CDN to achieve a high hit ratio but performs poorly on edge caching due to the fact that the local request by users may be different than the nation-wide distribution.}
    
    \item \emna{\textit{Least Recently Used (LRU):}  This conventional caching scheme can be considered as a baseline that has been used for a long time in networking systems. LRU stores the time of the last access of each content and when the memory capacity of the server is insufficient, it replaces the most idle video that was not requested for a long time, by a new content. \khan{The problem of LRU is that it gives priority to some unpopular content just because they were recently requested \cite{LRU1}.  LRU has multiple variants including the three Segmented Least Recently (S3-LRU), which divides the cache  into three segments where the most requested videos are stored in the highest segments and least requested are saved in the last segment. If a new content is requested, it is placed in the head of the list while removing the tail of the last segment. If it already exists in the cache, it is considered as the least recently requested and all the others are shifted downwards \cite{S3LRU}.}}
    
    \item \emna{\textit{Least Frequently Used (LFU):} This caching strategy relies on the number of requests per video to judge the popularity of content. In this way, the content with the lowest number of requests is evicted to create a room for the new published video. \khan{However, LFU suffers from gradual performance degradation as videos with historical high number of requests can remain a long time in the cache even if they are no longer accessed \cite{li2013popularity}. This problem is solved by defining a time window for observing the frequency of requests as done in the Jumping Window Least Frequently Used (JW-LFU) and the Sliding Window Least Frequently Used (SW-LFU).}}
    
    \item \emna{\textit{First In First Out (FIFO):} This scheme caches the videos according to the order of their first request. It means when a new video is requested for the first time, it occupies the head of the list.  The FIFO was later combined with LFU (FIFO-LFU) to take into consideration the popularity of the content \cite{FIFO_LFU}.}
    
    \item \emna{\textit{Least Recently Frequency Used (LRFU):} This caching approach takes into consideration both frequency of requests and recentness, as adopted in \cite{LRFU}.}
    
\end{itemize}

\emna{
The aforementioned traditional caching schemes are widely used in edge computing owing to their low complexity and simplicity of deployment. However, these approaches ignored the dynamic of viewers, and their preferences and behaviors. Therefore, due to the limitation of edge bandwidth and memory, better caching schemes that take into consideration the viewership and their capacities and make videos available at peak hours beforehand using prediction techniques should be designed.
}

\begin{figure}[h]
    \centering
    \includegraphics[width=0.9\columnwidth]{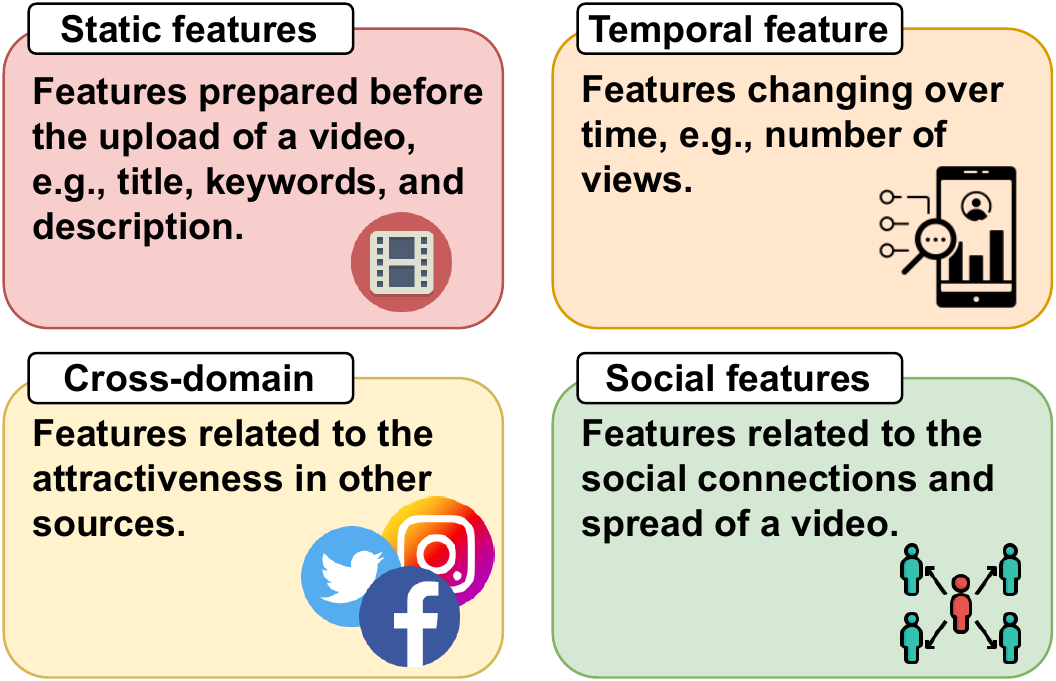}
    \caption{Popularity prediction features.}
    \label{context_popularity}
\end{figure}

\subsubsection{Human Behavior-Aware Strategies} This type of strategies is based on popularity prediction and the learning of the viewers’ preferences through either AI or mathematical studies. However, prediction techniques can add latency overheads and computational costs to ensure satisfactory performance and take accurate caching decisions. Therefore, the designed algorithm should be quick, provide accurate decisions, and scalable to handle high number of requests and large library of content. Finally, the prediction should be based on the preferences and behaviors of in-proximity viewers. Specifically, the contextual and social information, the content’ demands, the users’ interests, and the geographical location of viewers should be the input of the prediction model. These data is processed to get spatial, temporal and social insights and identify the viewing pattern. In this way, the designed caching strategy  will be able to predict the videos that will receive higher attention locally or globally, and accordingly plan for content storage, replication, eviction and select the most adequate MEC server for content storage. The input features can be classified into four groups: static, temporal, cross-domain and social features as we can see in Figure \ref{context_popularity}.

\emna{
\begin{itemize}
    \item \textbf{Static features:} refer to the parameters that are prepared before publishing the content. These features include the video characteristics that give hints about the quality of the video such as the duration, publication date, video and audio standard, music style, etc. The visual features such as the images at each frame and the text features such as the category, the title, the keywords, and the description can contribute to increase the popularity of the video and enhance its visibility in the search engines.   
    \item \textbf{Temporal features:} refer to the data that change over time such as the channel features including the number of rates, views, subscribers, comments and shares for different published videos and the video features such as the video age, the watch time, and the subscribers gained from this content. Additionally, the previous requests of viewers can give insights about their future requests for other content.
    \item \textbf{Cross-domain features:} refer to the external sources such as the reputation of the content creator and the propagation of the content via other social medias or video platforms.
    \item \textbf{Social features:} includes the relationship between users, their followers and followees, and their social interaction which can help to share and forward the content to more viewers.
\end{itemize}
}

\emna{	
A wise selection of the features fed as an input to the prediction model is very important to improve the accuracy, reduce the complexity and computation time of the decisions, and remove the redundancy of attributes. Next, we will discuss the popularity-aware MEC caching approaches that rely on the described features. These approaches can be classified into two groups: a single domain and cross-domain strategies. The single domain covers the approaches trusting only the features related to the video and its broadcasting platform such as:
}

\begin{itemize}
    \item 	\textit{Popularity evolution of a content:}  The high correlation between the past popularity of a videos (e.g., number of views, shares, and requests) can indicate its trend for the future. In \cite{ahlehagh_2014}, two caching policies i.e., P-UPP and R-UPP are proposed, which are based on the popularity of content. In P-UPP (Proactive User Preference Profile) scheme, videos that are Most Likely to be Requested (MLR) by active users of the cell are pre-fetched. \khan{This is different from the traditional scheme such as MPV, which accounts for the nation-wide popularity of the video content. The local popularity of a video $v_i$ in a pool of videos $V$ is calculated using the equation:} 
    
    \begin{equation}
        \khan{p(v_{i,j}) = \frac{p_{vc_j} (v_i)}{\sum_{i=1}^{|V|}{(v_i)}}}
    \end{equation}
    \khan{where the denominator of the equation represents the probabilities of all videos in the category $vc_j$}.
    A drawback of this scheme is that if the cell is highly dynamic, the users may leave or join frequently which will change the MLR set of videos, thus increasing the computation. 
    To reduce the computation complexity caused by cell dynamics, authors proposed to use a threshold value for cache hit ratio to decide before pre-fetching/replacing content. In R-UPP (Reactive User Preference Profile), a video is fetched upon user request, however if the cache is full, the video that is "Least Likely Requested" (LLR) is replaced. LLR is calculated using the probability of requesting a video.  \khan{In real-world, the demand of a video content is always associated with the user behavior i.e., it is the user preference that makes a video more popular. Hence, caching schemes based on the user context/behavior is proposed in \cite{Zeng2019}.} The authors proposed a smart caching scheme using knapsack problem to find context-aware content popularity to maximise video hit rate within a limited time. \khan{As the user preference may change based on time and/or location, in \cite{Zeng2019}, the spatio-temporal context is constructed from the user's access time and spatial characteristics to mimic video popularity.} 
    In \cite{gharaibeh_2016}, a collaborative online caching algorithm is proposed that minimizes the sum of User Attrition (UA) cost and caching cost for each content. The problem is formulated as Integer Linear Program (ILP). In the online settings, the user requests are revealed one by one to the edge server and the server should take decision before next request to provide the content from local cache (if available), or fetch from neighboring edge servers or from the origin server.
    Authors in \cite{baccour_2019} proposed Proactive Cache Policy (PcP) i.e., a popularity-aware proactive video chunks caching based on chunk popularity instead of the whole video popularity. \khan{This is particularly useful as normally the users would not watch the whole video but rather watches the first few chunks. Hence, this work proposes to cache the popular chunks rather than the entire video of long duration.}
    \item 	\textit{Metadata of the content:} Some videos are newly published and has not indication about the past. Hence, the content data such as subscribers, title, and frames can be considered to predict the popularity. Authors in \cite{metadata_popularity1} proposed an online caching strategy that uses the mixed-integer linear program (MILP) to load the MEC cache with most popular videos during the valley hours, based on the content metadata such as  keywords and description. Authors in \cite{metadata_popularity2} designed a deep learning approach that handles the popularity of new published videos. The input data of the model are extracted from the video raw data. Then, the popularity is defined by studying the similarity of the extracted features to the old videos.
    \item \textit{Social dynamics:} the interaction between users (e.g., friendship, likes, and shares) are very insightful for the watching trend of videos within a small edge area. The authors in \cite{social_popularity} used the susceptible-infected-recovery (SIR) model to study the social propagation of videos. More specifically, the SIR model defines three states: susceptible, informed and refractory. In the first state, the viewers are notified of the new published content.  During the informed state, the users discover the video and decide whether to share it to the followers or not. If the content is shared, more viewers can see it, if not the initial user is considered in refractory state. The popularity of a video can be examined through the probability to spread the content, which is also called social connections.
\end{itemize}

\emna{
The cross-domain prediction models use features from external sources (e.g., attractiveness of the video on other sources) to increase the accuracy of the decision. In \cite{Tran2020}, authors improve the popularity-aware video caching by jointly considering the popularity and attractiveness of the video stream. The authors argue that despite the high popularity of the videos, user may decide to finish the videos after a short duration of viewing (i.e 15 seconds). This duration is known as retention rate. Thus, simple heuristic algorithm is developed to cache videos with high popularity as well as low retention rates.
}

\subsubsection{Cooperative Content Caching}
\khan{In cooperative caching, multiple MEC servers collaborate with each other to serve the network-wide users' requests. Thus, a single MEC does not need to cache all videos. In fact, different MECs serve the requested videos to any user in the network to minimize the CDN cost and access delay, while maximizing the cache hit ratio. Several works propose cooperative caching in multi-server MEC environment.} For instance, in \cite{bilal_2019} authors proposed to fetch content from neighboring servers using high rate X2 interfaces (instead of typically used S1 interface), when the requested bitrate of the video is not locally available. The scheme aims at improve bandwidth utilization of backhaul links. \khan{Despite the fact that cooperative caching can help to efficiently utilize the network resources, storing multiple bit rates of the same video in overlapping regions is still a challenge. Authors in \cite{Qu2020} explain the trade-off between caching for high bitrate videos and caching for diversity videos.} \cite{Qu2020} propose an optimum caching algorithm to store multi-bitrates of videos at the edge servers aiming at maximising the user perceived QoE. \khan{The authors adopt the proactive caching policy to update edge caches when the system is idle (i.e., least busy).}
\khan{Video content has huge storage size that makes caching a large number of video files at the edge server challenging; hence, efficient cache replacement strategies are required. To cope with this problem, authors in \cite{Su2017} proposed to improve the performance of mobile video delivery through caching layered videos i.e., instead of caching the entire videos at the edge server, only layers/descriptions of videos can be stored at the edge. Thus when a user request is generated, the edge server can send the cached layer to the users, while in the meanwhile the rest of the video is fetched form the remote/origin server.}  
\khan{Similar to cooperation among MEC servers, network users can also implement cooperative strategies to serve contents to each other.} A relevant study on users' cooperation in vehicular network is found in \cite{Vigneri2019}. The work proposes a low-cost video caching strategy for mobile devices for delay-tolerant video applications and provided simulation evidences to claim the efficacy. While caching most popular videos at the edge servers significantly improves the performance of video delivery system, there is still a tradeoff between caching high bit rate videos versus caching high diversity videos.

Video retrieval occurs too frequently as compared to cache placement. Hence to enhance the efficiency of edge based caching, the caching policy shall consider the two different time scaled of cache placement and video retrieval. In \cite{Zhang2020a}, authors proposed a caching policy that jointly considers the long-term cache placement and short-term cache retrieval in coordinated multi-server system to reduce service delay and cache hit ratio.
In \cite{Cheng2020}, authors propose a hybrid caching scheme by combining edge cache sponsoring (ECS) and cellular data sponsoring (CDS). In CDS, the content provider serve free content to the user at the cost of playing advertisements to generate revenue. The user is allowed to select any scheme with the aim to achieve total maximum revenue for both ECS and CDS via cooperation or for an individual provider via competition.

To avoid caching a large number of multiple bitrate versions of the same videos at the edge server, a single high bitrate version of a videos is stored and then transcoded to lower bitrate when lower bitrates are requested by users. However, when the number of user requests increases, the simultaneous transcoding tasks can exhaust edge computation resources. To efficiently uses caching and transcoding resources at the edge server, authors in \cite{Kumar2020} proposed a RAN-aware adaptive caching scheme, which employs network information acquired from the RAN to estimate the probability distribution of the user requests and uses it to jointly decide video bitrate selection for caching and transcoding. In \cite{Li2020a}, authors jointly optimize caching, transcoding and retrieval in an energy efficient manner. The probability of uncertain user requests is inferred via historical data to propose a proactive caching policy. The work proposes that edge server can use backhaul links to retrieve requested but not locally available from other neighboring servers. 

\khan{Few works consider financial metrics (e.g., revenue etc) to design cooperative caching schemes. For instance, in \cite{gharaibeh_2016}, a collaborative online caching algorithm is proposed that minimizes the sum of User Attrition (UA) cost and caching cost for each content. The problem is formulated as Integer Linear Program (ILP). In the online settings, the user requests are revealed one by one to the edge server and the server should take decision before next request to provide the content from local cache (if available), or fetch from neighboring edge servers or from the origin server.}


\subsection{Edge-based Processing}
Cloud computing has been preferred for computational extensive tasks. However, it has several shortcomings in addition to the common issue i.e., delay constraints as discussed earlier. First, when transcoding a video stream at cloud, multiple versions are created. The number of such transcoded bit rate versions increases the traffic traversing the core network. Second, the cloud-based transcoding strategy is usually fixed and does not adapt to the dynamic changes in viewership. Hence, edge-based processing of video content improves the network resources utilization by sending only the higher bit rate version to the edge server and then the edge server transcodes videos locally as per local viewership. However, edge servers have limited computational resources as discussed earlier and hence must optimize computational tasks to improve the system's capacity. This section presents an overview of research efforts in MEC based video content processing.

\subsubsection{Joint Caching and Processing}
The edge server provides caching and computation resources in close proximity to the end devices. In video streaming applications, mobile devices may need to utilize both resources i.e., fetching cached video content from the edge server and offload computation tasks such as video transcoding to the edge server. For efficient utilization of both resources, the joint optimization of caching and processing functions can improve the network performance to achieve several benefits. In this section, we provide research efforts on joint optimization of caching and processing resource allocation. \par
The collaboration among multiple MEC servers connected via backhaul links is proposed in \cite{tran_2017}, i.e., for each new video request, servers collaborate with each other to jointly cache and transcode videos for each other. The idea is that any MEC server in the delivery path (from video origin) to the home MEC server of the requesting user can transcode the video. A major benefit of this scheme is that a server does not need to cache different bitrate versions of the same video, enabling efficient storage across the edge network.
\khan{Multi-MEC cooperation allows to efficiently utilize overall edge resources in a network. To improve the utilization of a single edge server's resources, end-devices must cooperate to access edge servers. One way to achieve this is efficient scheduling of tasks at the edge server.} In \cite{Yan2019}, authors propose an optimal scheduling strategy for video streaming in best-effort HTTP DASH-based video delivery, leveraging joint coordination among mobile users. \khan{Example of joint caching and processing using edge-device cooperation is in \cite{Yang2018a}}. In this work, the authors investigated edge-based processing in VR application. While the VR devices cache most of the components locally, the MEC delivers the components not cached at the VR device. The components are cached and processed at the edge server to reduce computation delay. In order to cope with high communication delay of the component delivery, a task scheduling strategy is proposed. \khan{Joint caching and processing has a great significance in adaptive video streaming in which the video bit rate is automatically adjusted (or selected) based on the user device capability (e.g., channel, buffers capacity etc). Instead of storing all possible bit rates of a video, MEC servers decides which bit rates of a popular video are stored and which bit rates are to be generated upon request.} For instance, authors in \cite{Tran2019} proposed joint caching and processing to implement adaptive bitrate video streaming by formulating it as integer linear program to minimize the the latency of video retrieval. The authors used ML technique to find the video popularity for caching. \khan{As caching large number of videos incurs storage cost, caching/processing decision also involves revenue as an important consideration.} As a example. authors in \cite{Hao2019} studied edge based joint caching and processing to maximise the profit to video service providers. The joint scheme is formulated as a binary optimization problem and solved using multi-armed bandits (MAB) problem and the cache is updated in real time as per users requests. \khan{Another interesting approach to solve the joint caching and processing problem is to use a more direct and realistic metric such as the "number of user requests". The work in \cite{Tang2019} addressed this issue by proposing a joint caching and processing scheme which aims at the maximising the number of service requests that can be served by each base station using a Stackelberg game}. The edge server predicts the requests of each BS's users and defines the caching/computing price that maximises its utility. The BS then compete with each other to maximise the availed resources from the MEC at fixed price. \khan{A similar approach to \cite{Tang2019} is proposed in \cite{Zhang2020b} with additionally considering the user association.} The problem is formulated as a mixed integer programming (MIP) to minimize the video retrieval latency in ultra dense heterogeneous networks. 

In \cite{Baccour2020}, authors propose a joint video caching and transcoding for VoD streaming called as "proactive caching and chunk processing (PCCP)". First video chunks of popular videos based on the user viewing patterns are proactively fetched and stored in neighboring edge servers such that none are replicated. When user requests a video, the respective chunks corresponding to the same video are collected from the neighboring servers and served to the user. A similar approach is used in \cite{Bilal2019} to fetch only high bitrate versions of a video from the origin/CDN server and transocode to lower bitrate versions at the edge servers. Different edge servers in the neighborhood can collaborate to transcode and share videos with each other via X2 backhaul interface. In \cite{Zhang2020}, authors propose a joint caching and offloading scheme to offload duplicate computation tasks and the requested data content to the edge servers. \khan{By jointly optimizing the caching and offloading decision, the scheme minimizes the latency while satisfying the energy consumption of mobile devices. The work uses genetic algorithm to implement an online-learning without requiring future information.}

\subsubsection{Joint Radio, Caching and Processing}
Resource allocation in edge computing can be further improved by considering the user's wireless channel quality in the resource allocation problem. Several works propose the joint optimization of the edge resource allocation and the user's communication resources to meet the QoS requirement. Some representative works are listed \cite{Dang2019, Sun2019, Luo2019, Wang2020h, Rezvani2018, Sun2020}.

In \cite{Dang2019}, authors proposed a joint radio, caching and processing scheme for MEC-assisted VR applications. The edge server proactively caches some parts of the videos and process these videos. The caching and processing resource allocation at the edge is jointly optimized with the transmission rate constraint of end VR devices to maximize average delay, while guaranteeing transmission rates, cache size, energy consumption and front-haul capacity. In \cite{Sun2019}, MEC-assisted VR delivery is proposed in which some parts of the videos are cached at the edge whereas others are stored at the VR device. The algorithm jointly optimize caching and computing resources to determine the parts of videos (more specifically called Field of Views or FOVs) to be cached and processed at the mobile device and which ones to offload to the edge server under three constraints (i.e., cache size, power consumption and latency).
\khan{Due to the more dynamic and complex nature of the problem, advanced techniques based on DL are being proposed. For instance,} in \cite{Luo2019}, the joint caching, transcoding and transmission of videos is formulated as Markov Decision Process (MDP) and solved using Deep Reinforcement Learning (DRL) to improve user experience in adaptive video streaming. In \cite{Wang2020h}, authors consider joint optimization of caching, processing and radio resources to maximize system revenue. In \cite{Rezvani2018}, authors propose video caching and transcoding scheme in heterogeneous virtual MEC networks. The scheme jointly optimize caching and transcoding with the network radio conditions during the cache placement and delivery phase respectively. In \cite{Sun2020}, authors propose to perform the pre-processing of the videos at edge servers using a lightweight Deep Neural Network (DNN) model, and upload the results to cloud nodes for further analysis to produce a complete video analytic solution. The aforementioned state-of-the-art discussed in this section is summarized in \khan{Table \ref{tab:mec_stream}.}

\begin{table*}[htbp]
    \centering
    \renewcommand{\arraystretch}{0.99}
    \caption{State-of-the-art of MEC in Video Streaming.}
    \label{tab:mec_stream}
    \colorbox{myyellow!0}{
    \resizebox{0.99\textwidth}{!}{
    \input{tables/mec_stream_soa}

    }}
\end{table*}

\subsection{Lessons learned and challenges}
\eb{
The main objective of deploying video streaming applications in the mobile edge networks is to minimize the perceived delay, the network cost and the energy consumption, and to maximize different resources utilities (e.g., computation, memory, and bandwidth) and the users’ QoE. Hence, most of the recent works in the literature focus on establishing a joint optimization to balance different purposes. 
The MEC strategies are classified into centralized and collaborative video caching and processing. The centralized approach relies typically on a base station or eNodeB to deploy different tasks. In general, this strategy is simpler and the optimal resource decisions are less complex. On the other hand, the joint strategies involve multiple MEC entities that collaborate to offload the streams, while respecting the objective constraints.}

More specifically, when the users' demand increases, MEC servers can collaborate to serve content to its neighboring servers. The collaboration among MEC servers becomes essential when storing all requested video content locally is not possible due to limited storage capacity. 
Typically, the caching and processing of content is considered as a joint problem in various works to improve the overall network performance. Moreover, due to its strong impact on the offloading capability, the channel quality is also considered in the joint optimization problem.
\eb{
In this context, previous works proved that the optimal solution is NP-hard and approximation algorithms have to be designed.
}

Due to the large size of videos and increasing use of video-based services, traditional cloud-based caching schemes such as MPV, LRU and LFU schemes are inefficient to meet the network demand. MEC-based video caching can benefit from machine learning to implement proactive content caching at the network edge to reduce latency and improving the utilization of limited caching capacity.


\section{Device-to-Device (D2D) Cooperation in MEC Systems}
\label{sec:mec_d2d}
Over the last decade, a significant increase in the computational power in the end devices has been observed. These resource-rich devices can collaborate with each other via D2D communication to achieve network-wide performance gains. In the MEC system, mobile devices can implement D2D communication to achieve several benefits: First, to alleviate the computation load on the edge server, particularly if the edge server has limited resources or high computation load by offloading tasks to mobile devices. Second, devices can collaborate to exploit and reduce communication latency by exploiting the high speed D2D communication. Third, due to the short range communication, the device's energy consumption can be significantly reduced. Fourth, the spectrum utilization of cellular network can be multiplexed to improve system's overall capacity. \par

\begin{figure}
    \centering
    \includegraphics[width=0.7\columnwidth]{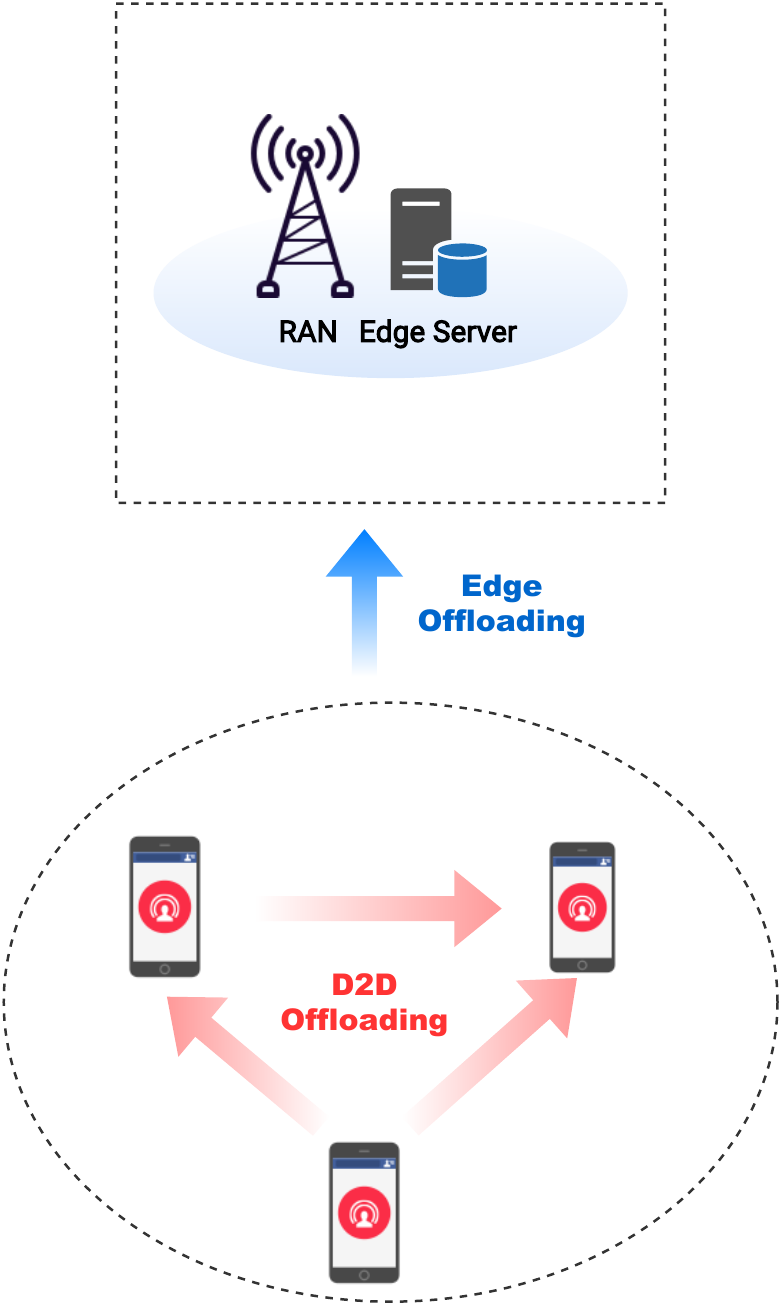}
    \caption{D2D Cooperation in MEC System.}
    \label{fig:d2d_mec}
\end{figure}

Both D2D and MEC systems aim to benefit from the proximity of the devices to achieve performance gains. In D2D networks, a device connects with another nearby device to communicate at smaller distance and thus achieve high data rates and less latency. In MEC systems, devices can benefit from the closely located edge servers to fetch content and offload computations thus achieving low latency and computational gains. Thus, the integration of both systems will intuitively allow network users to benefit from both technologies in a range of applications. Figure \ref{fig:d2d_mec} illustrates D2D offloading in MEC systems.

\subsection{Relevant Works} \label{subsec:rel_works}
D2D communication has been used in wireless networks to gain several benefits such as energy saving, less bandwidth utilization and enhanced quality of experience \cite{yaacoub_2012, asif_2019, asif_2019b}. For instance, \cite{yaacoub_2012} proposes a generic framework for D2D communication to improve video quality and reduce energy consumption and bandwidth utilization. In \cite{asif_2019b}, authors propose Wi-Fi Direct based D2D scheme to scale the network size and increase throughput. \par

Recently, D2D-enabled MEC systems have been proposed \cite{li_2014, pu_2016, hu_2017, yuan_2019, he_2019, qiao_2019, gope_2019, park_2019, zhang2020mec, kim_2020}. In D2D-enabled MEC systems, mobile devices can not only exploit the communication resources over D2D links but they can also benefit from the computational resources of under-utilized devices. Every D2D enabled device in the network is a computational resource and thus can aid to improve the overall computational capability of the system by realising cooperation among devices and edge server. Such kind of D2D-MEC cooperation can be extremely useful. \par

Several works have recently proposed the integration of D2D communication in MEC systems to achieve performance gains. In \cite{Li2014}, authors proposed that D2D communication can be used by mobile users to find alternative resources when accessing the cloud server encounters long delays due to intermittent wireless connectivity. D2D-fogging is proposed in \cite{Pu2016}, a concept similar to standard edge computing architecture with D2D communication enabled among user devices. The D2D devices collaborate with each other to share their computation resources controlled by the base station (associated edge server). The tasks are offloaded online, aiming at minimizing the time-average energy consumption. In \cite{Qiao2019} authors proposed a D2D-ECN (Edge Computing Network) framework for computation offloading. Using D2D-ECN, computation intensive devices can offload tasks to resource rich devices. Q-learning has been used to perform optimal resource allocation in a point-to-point offloading system. In \cite{Wu2019}, authors proposed D2D-MEC system in which devices offload their computation to nearby idle devices (helpers). The task offloading is jointly optimized with communication and energy beamforming to maximize the sum-computation rates of users. In \cite{He2019} authors used D2D communication in MEC system to maximize the number of devices supported by the system with communication and computation constraints. A mixed integer programming (MIP) formulation is presented for the D2D-MEC system and solved by decomposing the MIP problem into two sub problem. The simulation results reveal significance of the proposed D2D-MEC system in cellular networks. In \cite{Yuan2019}, authors highlighted the use of D2D collaboration in MEC system to improve the overall system capacity. Computing in D2D network is more complicated when it comes to resource and topology management. The work in \cite{Kim2020} presents resource management in D2D-MEC system such as link selection and sub-channel allocation, transmit beamforming, transmit power and receiver combiner.

In \cite{dogga_2019}, authors investigated experimentally the feasibility of video transcoding at the user devices. The authors transcoded videos of short duration on mobile phones to measure the transcoding time. In \cite{he_2019}, authors propose a D2D-MEC system to improve the computation capacity of the network i.e., to maximize the number of devices supported by the cellular networks. In this scheme, a device can offload its tasks to an edge server and a nearby D2D device. In \cite{wang_2017} a joint communication and computation optimization scheme for MEC is proposed that leverages on D2D communication. The nodes use collaborative beamforming to wirelessly charge other devices and offloading computational tasks to idle devices. In \cite{xing_2019}, authors studied D2D-enabled multi-MEC system in which a user offload computational tasks to multiple local users using time division multiple access (TDMA) transmission to achieve latency reduction. The timeslots are divided into three phases. In the first phase, the task is offloaded to the helper nodes. In the second phase, the helper node executes the task. The computation results are downloaded in the third phase.

Quantile-based CSMA \cite{kim2014spatial} and  ITLinQ \cite{naderializadeh2014itlinq}  have been proposed  to improve the network throughput for video transmission in  static  networks without consideration for dynamic user mobility. In \cite{goebbels2008enhancements, dehghan2015complexity}, the authors proposed the caching at the network edge to reduce the congestion and the delay transmission for video content.  The caching technique is also proposed in  \cite{ golrezaei2014base, ji2017order} where the D2D communication is used to create  a direct communication between nearby mobile users. These caching solutions are effective for  stored video but they introduce higher throughput, delay, and jitter in case of live streaming. 
In \cite{wu2013flashlinq}, the authors proposed FlashLinQ based on CSMA protocol to establish a D2D link. The links are classified according to the link's priorities where the links with higher-priority do not suffer from significant interference with lower priority links. The algorithm uses the priority for each link to schedule its activation time. FlashLinQ has a good performance in terms of latency; however, the video quality-aware mechanism was not  considered in this work and therefore its suitability for D2D on-demand video streaming remains open. 
D2D communication has been proposed for video streaming in \cite {abbasi2018multimedia}. The proposed model  improves the resource utilization in 5G by using a scheduling algorithm for
effectively sharing the multimedia content using D2D communication. However, this algorithm studied only the case for one user and one video without considering the case for a multi-user with multi video sources.
In \cite{ren2019video}, D2D multicast communications for live streaming video is investigated.  The proposed scheme uses the frame priority (FP) to improve the QoS perceived by users and the users' satisfaction with the video quality.  The proposed model considers the encoding characteristics of video streaming and users' feedback to ensure that the frames to be retransmitted are valuable for decoding.

Table \ref{tab:d2d_mec} summarizes the recent research studies on D2D-enabled MEC systems.

\begin{table*}[htbp]
    \centering
    \caption{Device-to-Device (D2D) Cooperation in MEC Systems.}
    \label{tab:d2d_mec}
    \colorbox{myyellow!0}{
    \input{tables/d2d_mec}
    }
\end{table*}

\subsection{Lessons learned and challenges}
Device-to-device collaboration for communication is an efficient technique and is widely proposed in wireless networks. D2D standard protocols also exists for direct communication among mobile users. D2D communication is used for improved network coverage, higher throughput and reduced energy consumption. In MEC, D2D has been proposed to improve the offloading capacity of the edge server. Particularly, devices can offload computational tasks to helper devices instead of the edge server. Such D2D offloading can be efficient when the total delay (sum of computation of communication delays) for local processing or edge-based processing is higher then D2D processing. D2D-offloading can also be used on the downlink i.e., when the edge server is overloaded, the server can offload user's tasks to the resource-rich helper devices.

\eb{
Incentive mechanisms need also to be envisaged to systematize the revenue between broadcasters, MEC infrastructure and D2D-participants.  Furthermore, the security and privacy aspects of the D2D offloading for video crowdsourcing applications are not well considered in the literature and they are still the major limitation for such an approach. In this context, blockchain-based techniques have been proposed as a strong tool to handle an incentive and trusted environment for many applications. However, it is rarely studied for edge-assisted video streaming systems. We believe that blockchain combined with distributed and cooperative network can guarantee data integrity and effectively realize a trusted topology, particularly in 5G networks. Moreover, since blockchain provides limited data storage, efforts should focus on  establishing a balance between off-chain and on-chain strategies.
}

\section{Intelligent Edge for Video Streaming}
\label{sec:mec_ai}

Both Machine Learning (ML) and edge computing are making profound impact in several domains, however when they are combined, they can bring a more intriguing user experience. The intersection of both is being considered as a suitable platform for vertical applications e.g., automotive (e.g., autonomous cars), healthcare \cite{chkirbene2020smart} (e.g., remote surgeries, real-time assistance),\zina{security \cite{chkirbene2020weighted} (e.g., intrusion detection)}, manufacturing (e.g., predictive maintenance), retail (e.g., VR-enabled online shopping experience and personalized suggestions) and connected homes (e.g., temperature control, smart smart doorbells, access control, and smart lighting). Most of the video streaming services that benefit from edge computing are using machine learning or have the potential to benefit from it.

Deep learning (DL) is a sub-domain of machine learning which has been successful in solving highly complex problems. However, to meet the computational requirements, DL applications typically leverage cloud computing. \asif{An associated drawback of running deep learning on the cloud is the that data might need to be transferred to the remote cloud server \cite{redondi_2016}. If such data is not available at the cloud (e.g., data coming from visual sensors such as IoT video sensors) \cite{chen_2020}, this "can result in network congestion and delays due to the long-distance or multi-hop transmissions" \cite{cao_2018}. Two illustrative examples of scenarios which involves transmission of videos to a server for processing include object/event detection (e.g., automatic fire detection, crime detection) using videos from surveillance cameras, and driving assistance in autonomous driving (e.g., sending captured videos from preceding vehicles to the server).
To cope with this challenge, edge computing comes as a suitable choice. By processing such data at edge servers, the latency is reduced and most of the data required to run the DL models is either available at the edge or at closely located end devices, hence improving bandwidth utilization.} \par


\emna{
One of the broad applications of DL for multimedia streaming is the video analytics for surveillance systems, which was introduced in the previous section \ref{Analytics}. In fact, we showed that some monitored areas are very critical and need continuous surveillance, such as military borders. In this scenario, video frames (or video streaming) are sent instantaneously to remote servers for real-time  object/accident detection at every small interval, which incurs huge data transfer. This type of applications has also stringent latency constraint, as it requires a prompt intervention, if an accident is detected. }
\par
\emna{
Furthermore, for a better accuracy of the results, high-resolution videos having high data volumes are transmitted by the source to the computation units. The accuracy is also affected by the speed of feeding the data to the model, which requires bandwidth availability and efficient communication technologies to follow the requirements of the DL model. It is worth to mention that video surveillance cameras are highly pervasive. For example, 170 million cameras are in deployed in China roads only \cite{china}. These devices are generating a huge amount of data as previously described. It is reported that the data produced by IoT devices will witness a growth of 28\% by 2025 \cite{surveillance_bigData}, where 65\% of it is related to the surveillance cameras \cite{bigData}. Even-though the conventional wisdom resorts to the centralized cloud servers for analytics owing to their high computational capacity, offloading these bulk of data presents scalability issues, as the  access to the cloud can experience bandwidth bottleneck when the number of video sources increases. Moreover, due to the strict latency requirements, the cloud is no longer a feasible solution.
} 




\zina{
Recently, DRL techniques have been applied for video streaming and network selection in \cite{n1, n2, n3}. In \cite{n1}, a system that generates  adaptive bitrate (ABR) algorithms using reinforcement learning (RL) is proposed. This model trains a neural network model  to select bitrates for future video chunks based on observations collected by client video players. In \cite{n2}, the authors proposed QARC (Quality Aware Rate Control) algorithm to obtain a higher perceptual video quality with possible lower sending rate and transmission latency. QARC uses DRL algorithm to train a neural network for selecting future bitrates based on previously observed network status and past video frames. In \cite{n3}, a predictive panoramic video delivering based on DRL has been proposed. The LSTM model is used to predict the user’s field of view (FoV) in the next few seconds. These systems focus on the continuous prediction of video resolution and average video bitrate.
}

\begin{figure*}[!h]
\centering
\includegraphics[width=0.7\textwidth]{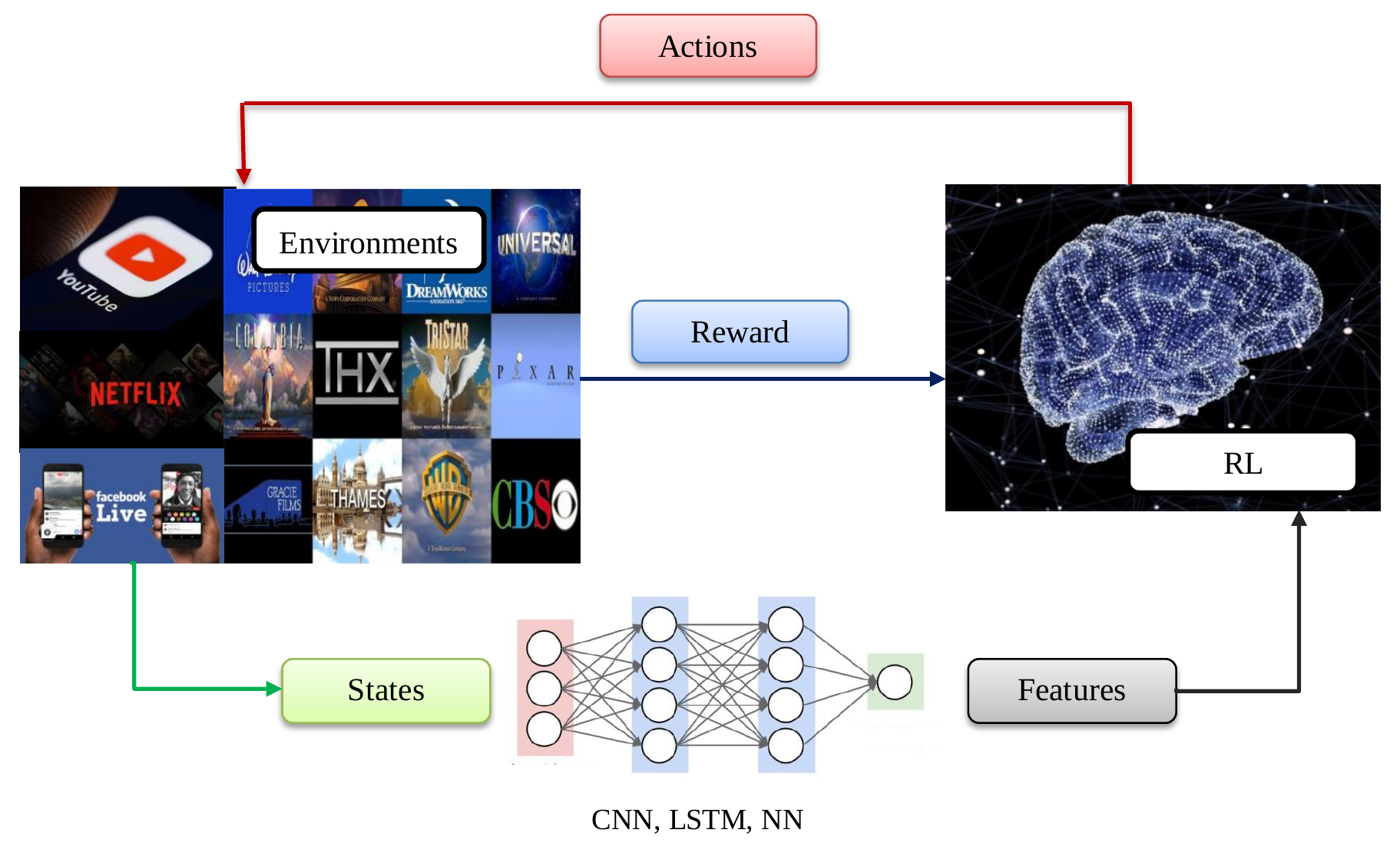}
\caption{DRL for video transmission.}
\label{fig1rl}
\end{figure*}

In the following, we provide a more detailed review of the research contributions on the use of ML/DL techniques in MEC systems for video streaming use cases. The section covers various areas of video streaming i.e., proactive caching, optimal computation offloading, adaptive streaming, video analytics and AR/VR streaming. A summary of these works is presented in Table \ref{tab:mec_ai}.

\begin{table*}[htbp]
    \centering
    \caption{Machine Learning in Edge Computing for Video Streaming.}
    \label{tab:mec_ai}
    \colorbox{myyellow!0}{
    \input{tables/mec_ai}

    }
\end{table*}

\subsection{Proactive Caching}
Machine learning is being widely proposed to implement proactive caching. Proactive caching is extremely useful when the user viewing patterns are not constant and resource allocation is required to meet service requirements. Machine learning thus brings new possibilities by implementing intelligent edge caching to capture hidden features from historical data or current network stats. For instance, ML techniques can be used to forecast video popularity or predict future user requests based on historical viewing patterns of users. ML assisted proactive caching schemes are investigated in \cite{li2013popularity, wang2017edge, tanzil2017adaptive, Tan2018, zhu2018deep, zhang2019toward}.
In \cite{li2013popularity}, authors integrated three ML techniques i.e., Auto Regressive Integrated Moving Average (ARIMA), Multiple Linear Regression (MLR), and k-Nearest Neighbor regression (kNN) to predict the users social patterns and used it for caching resource allocation. Advanced prediction techniques such as Long Short-Term Memory (LSTM) deep learning model can be used for more efficient prediction of video popularity. LSTM has been effectively used in \cite{zhang2019toward} to predict both long-term and short-term video content popularity. While historical user viewing data can provide a reasonable estimate of future behavior, other parameters can also be considered for effective caching policies. In \cite{zhu2018deep}, authors applied Deep Reinforcement Learning (DRL) to implement caching scheme by using user requests, network constraints, and external information. \khan{In the proposed DRL-based scheme, the inputs are the index of currently requested content, the number of requests for the cached content, and the time when the cached content was most recently requested. Using these inputs as "states", the DRL agent determines whether or not to cache the currently requested content. The reward function is the offloaded traffic in each request. The DRL model evaluated using the MovieLens dataset \cite{movielens_dataset} shows improved performance against traditional caching policies i.e., FIFO, LFU, and LRU.}
In \cite{tanzil2017adaptive}, a network-aware caching scheme is implemented using machine learning. The users' behavior and networks parameters such as cache size, bandwidth, and load are to build a neural network model to estimate video popularity. The estimated video popularity of video content is used in mixed integer programming to calculate cache size and content placement. Authors in \cite{Tan2018} developed a DQN model to implement a mobility-aware joint caching and computing design in vehicular networks. \khan{The model states include parameters i.e., available caches, available vehicles, contacts per time slot, and contact times for every vehicle. The agent performs the action i.e., decides which server and/or
vehicle is assigned to the requesting vehicle, and determines whether to cache the requested content or not. The reward function is cost of communication for data transmission.}

\subsection{Computation Offloading}
The computation offloading and resource allocation at edge servers is often solved as an optimization problem using integer programming or heuristics \cite{baccour_2019, baccour_2019b, baccour_2020}. However, recently deep learning models such as DRL and Deep Q-learning networks (DQN) are proposed to solve resource allocation in various networks. For instance, in \cite{wei2018joint}, DRL has been employed to jointly optimize caching, computation offloading policy, and radio resource allocation to minimize the average end-to-end delay. \khan{The DRL model in this work uses state parameters including number of requests, number of fog nodes, task sizes, computation requirement, popularity of requested content, and channel quality vector between devices and fog nodes. The action space includes multiple decisions such as the fog node to serve the user, whether to store the requested content or not, the allocated communications sub-channels and computational resources. The reward function is the weighted sum of time for computation and content delivery.} In \cite{huang2019deep}, the edge resource allocation problem considering multiple tasks offloaded to the edge server by mobile devices is investigated. The problem is formulated as mixed integer program and then solved using DQN. \khan{The system state includes all users’ offloading decisions and bandwidth allocations for all users. The action is an index that denotes the selection among two different neighboring states. The reward is chosen as 1 if the total offloading cost is decrease after taking an action and -1 if the cost is increased. If the offloading cost remains constant after an action, reward is 0. The total offloading cost is the sum of energy cost, computation cost, and delay cost.}

\subsection{Adaptive Streaming and QoS}
Adaptive streaming is the ability of the video delivery system in which the video quality is automatically adjusted according to the user's preference or network conditions. More specifically, the video bitrate is automatically adjusted as the user's channel quality degrades or improves, or the user requests a particular bitrate. Although mobile edge computing provides extra support to implement adaptive video streaming in video services, recent development in deep learning particularly deep reinforcement learning are being proposed to improve MEC based adaptive streaming.
In \cite{de2017qoe}, authors proposed a multi-stage learning system using Restricted Boltzmann Machine (RBM) to manage simultaneous video transmission which guarantees a minimum quality level for each user. In \cite{wang2019intelligent}, authors used DRL model to assign users to the most suitable edge server aiming at improving quality of service in video streaming. \khan{In this model, the states include resource usage, viewer scheduling information, and current viewer request. The agent's action is to assign a viewer's request to a server. When a new user's request is assigned to a server, it will increase the load on the system which impacts the QoE for the new users. Thus, the reward function is chosen as the opposite number of the overall penalty for the coming viewer.}
\khan{The aforementioned works on adaptive video streaming do not include mobility scenario such as vehicular users. To implemented MEC-assisted video streaming services in road scenarios, authors in \cite{yang2021} propose the use of DRL technique. The states of the DRL model include video quality level, queue length in the edge server, and average wireless channel gain. The wireless channel gain captures the mobility pattern of users. The agent uses the states to decide the action i.e., assigning the video quality levels to users. For each action, the reward is computed i.e., weighted sum of high quality level assigned to user, negative reward of quality level switching and the negative reward for queue length.}

\subsection{Video Analytics}
\khan{Video analytics are advanced services using streaming videos to extract real-time insights. There is a broad range of applications which requires real-time analytics in streaming services such as surveillance \cite{ren2018distributed}, autonomous vehicles, cognitive assistance \cite{liu2017new}, video summarization \cite{Wang2020c}, and filmography etc. However, such analytics are only useful when provided in real-time i.e., within strict delay deadlines. These advanced services rely on deep learning techniques which requires huge computational resources. MEC is being increasingly proposed as a promising solution to perform faster analytics and at lower data transmission delay.}
Authors in \cite{ren2018distributed} used CNN model for object detection in edge computing based video analytics for real-time surveillance and achieved accurate Region Of Interest (ROI) detection. \khan{The edge servers train local models on data collected from devices connected to it in a distributed way whereas the cloud server contains a global model which is shared with the edge servers.} In \cite{liu2017new}, CNN model is used for food object recognition for latency-sensitive application using edge computing services. \khan{The recognition latency and energy consumption of devices is reduced by distributing the model learning tasks among edge servers and the cloud server. The initial image pre-processing such as identification of blurred images is performed at devices using simple methods i.e., K-means algorithm. Then image segmentation i.e., extracting background and foreground object from the image is performed at the edge layer. Lastly, the DNN model i.e., CNN for object recognition is executed at the cloud.} In \cite{ran2018deepdecision}, a CNN model, Yolo \cite{yolo} is implemented on edge server to support computationally weak front-end devices by offloading their tasks to the edge to achieve a higher frame rate and accuracy. \khan{The scheme estimates the network conditions and the actual requirements of the application to decide the offloading strategy in CNN computing.} DeepCham, a DL-based object recognition model is proposed in \cite{li2016deepcham} using collaboration among users and edge server. Devices and edge servers collaborate to train the CNN model to improved accuracy of the model. Similarly, a lightweight CNN model is developed in \cite{nikouei2018smart} for resource-constrained edge server with reduced numbers of filter numbers in each layer. \par

Video summarizing on edge server and sending the video summary instead of the whole video to let the user understand the video content have many attractive use cases. User experience can be significantly improved if the video summary size is tailored according to the varying network bandwidth and user viewing behavior. Authors in \cite{Wang2020c} proposed latency-aware edge computing to achieve the adaptive summarization with improved bandwidth utilization and viewing experience. \khan{In \cite{galano2021}, authors applied DNN to automatically perform object recognition in videos. The proposed method improves the accuracy of the object recognition while meeting the frame capturing rate constraint.}

\subsection{AR/VR}
\khan{MEC is being used in AR/VR streaming services to accomplish DL tasks such as object recognition and features extraction at much higher rates. This involves relatively more complex tasks such as field of view (FOV) prediction in in $360^{\circ}$ using deep learning \cite{hou2018predictive, xu2018gaze, afzal2017characterization}}. The predicted FOV is used to determine the spatial region in video to be fetched from the VR content provider. Edge computing can help to improve such applications by reducing the latency as illustrated in \cite{ha2014towards}.
In Table \ref{tab:video_datasets}, we have provided a list of useful datasets of video streaming with brief descriptions and applications for prospective researchers.

\begin{sidewaystable*}
    \centering
    \caption{Datasets for Video Streaming Research.}
    \scriptsize
    \colorbox{myyellow!0}{
    \input{tables/video_datasets}
    }
    \label{tab:video_datasets}
\end{sidewaystable*}

\subsection{Lessons learned and challenges}
Machine learning techniques can be efficiently applied in mobile edge computing to gain several benefits. In video caching, ML techniques can predict the cache demand to allocate storage resources and prefetch video content before users request them. User requests can be accurately predicted using user's viewing patterns or video popularity on both short-term and long-term basis. LSTM and DRL techniques are relatively efficient for proactive video caching. For computation offloading and resource allocation on the edge server, DQN and DRL have been widely proposed. 
\asif{In addition to these, MEC also provides a platform to deploy ML-based services for several applications including video analytics, interactive videos, and content detection.} \khan{A detailed analysis of the ML-based techniques discussed in this section shows that MEC improves the application performance in several ways; (i) perform training and inference of DL models fully or partially, (ii) MEC can be used for pre-processing of huge amount of data, (iii) collaboration among edge servers and cloud improve system's capacity.}
\eb{Such applications (e.g., VR and AR.)  do not only require a successful classification of the objects, but also a high accuracy rate. However, the stringent requirements in terms of latency halt them from computing remote tasks. In other words, even 100 ms end-to-end latency of edge classification may no longer match the highly variable location of moving objects \cite{compression}. In this context, authors in \cite{sickness} expected a latency equal to 20 ms to avoid motion sickness in virtual reality applications. Compression techniques to minimize the data transmission latency are proved to be promising \cite{compression}. However, compression strategies should be revisited to establish a trade-off between data transmission and classification accuracy. Moreover, edge-assisted on-device inferences are envisaged to avoid remote transmissions.}

\section{Future Research Directions}
\label{sec:conclusions}

\subsection{Mobile Edge Computing in 5G/6G Networks}
The current 5G networks have the potential to provide a robust network for connecting people around the world. However, challenges persist as new services with more stringent requirements are introduced. The next generation of wireless communication known as 6G \cite{rappaport2019wireless, letaief2019roadmap} can meet these unprecedented requirements that 5G might not support. 
\eb{
In fact, when we deep dive on the roadmap of the 8K, VR, AR, and even eXtended Reality (XR), we can point that it not possible with 5G to longer support these volumetric videos. This is due to the 5G design and baseline KPIs that aim to reach an average ubiquitous capability equal to 100 Mbps downlink and 50 Mbps uplink \cite{6G_rev}. Whereas, the long term roadmap of eXtended reality videos is to be pervasive everywhere on a network offering multi-gigabytes to users. Moreover, current 360° 4K videos need 10 to 50 Mbps to be delivered to devices, meanwhile the next generation 8k videos require from 50 to 200 Mbps, which cannot be served by 5G. Also, looking to the futuristic types of videos, such as 6 Degrees of Freedom streamings, and holography presenting full immersive experience and requiring few Gbps to Tbps, 5G network is highly challenging. Aside from the incapacity to deliver a fast data rate for volumetric applications, 5G is not designed for ultra-precision streaming tasks, such as remote surgeries. On this basis, future works on 6G paradigm should think about enhanced Ultra-Reliable and Low-Latency Communications (URLLC) services, advanced massive machine type communications (mMTC) services, and eventually new network capabilities that allow to deploy pervasive streaming on IoT devices or to fuse physical and virtual technologies. More specifically, a new research area namely network slicing has emerged \cite{slicing}, which consists of dividing a single network connection into several virtual connections providing distinct portions of resources that serve different types of traffics. An interesting direction could be the enhancement of volumetric video streaming in 6G networks via resource slicing.}

Another major transformation from previous cellular networks generations that 6G will bring, is the use of artificial intelligence (AI) in the entire system architecture, from core to the edge of the network. This transformation will be an evolution towards "connect intelligence". 
\emna{
In the video streaming context, 6G networks can use AI to perform big data analytics of multimedia content to gain insights on network performance, channel conditions, viewers perspectives and preferences, and viewers' location prediction.  MEC plays a key role in 6G networks as it can operate as an intermediate layer providing localized and low-latency data processing for real-time applications. More specifically, the high capacities of centralized cloud servers offer sophisticated video and metadata analysis, while sacrificing in terms of transmission overhead. Using the MEC allows to divide the data analytics into tow phases. First, fast content caching and processing can be done at the edge to perform critical tasks, owing to the proximity of MEC servers to viewers. As a second step, a deeper and more efficient data processing is carried out in the remote servers, at the expense of higher delays. Resource allocation decisions for live videos, cache placement of content, video recommendation, and advertising process can be cited as relevant use cases \cite{letaief2019roadmap}. 
\eb{
However, given the limited  resources in MEC networks, computing the AI training/inferences in MEC devices may be infeasible,  particularly  when  the  task  requires  high  computational  load, e.g.,  Deep  Neural  Networks.  A  promising  solution  is to  adopt  pervasive  computing,  where different data storages and  processing  capacities  existing  everywhere  and  including distributed  edge  servers and  users  devices cooperate  to  accomplish  AI  tasks  that  need  large  memory and  intensive  computation.  This  marriage  of  pervasive  computing in MEC networks and  AI  has  given  rise  to  a  new  research  area,  namely “Pervasive  AI”,  which  garnered  considerable  attention  from both  academia  and  industry \cite{pervasive_AI}.  Formally,  pervasive  AI  focuses on how to intelligently distribute the inference or the training of  the  AI  model  across  edge devices,  to  minimize  the  latency, and  improve  privacy  and  scalability. Research of this paradigm is still in its infancy and needs further investigation to improve the deployment of AI for video streaming tasks in mobile edge computing networks.}
}

\subsection{Resource Migration with User Mobility}
While the purpose of edge computing is to bring storage and computation closer to the user, mobile users often move around from one edge server to another one and resource migration is required to provide high quality of experience. Consider the example of a tourist moving around a city with interactive glasses through which he receives video content about the places he sees through it. While the video content is always served through the nearest edge server to the respective place he is currently viewing, the tourist might move away (e.g. sitting in a tourist bus) while he still wishes to continue watching the video of the previous place. In such a scenario, the previous edge server needs to migrate the content and computation to the new server closer to to the user's current location. Another use case when resource migration might be required across edge servers, is to emergency treatment provided to a critical patient in an ambulance moving towards the hospital. The ambulance on-board facilities are connected via 5G links to the healthcare provider central data and computing resources. As the ambulance is traveling, it changes association to different edge servers and consequently the resource migration is required. The resource migration in the above two use-cases have different latency constraints i.e., the second case have more stringent latency constraint. The current state-of-the-art in resource migration is not sufficient and more investigation is required to provide the required QoS in different use cases involving resource migration in edge networks.

\subsection{Live Video Analytics for Drones}
Drones also termed as Unmanned Aerial Vehicles (UAVs) are becoming popular and affordable and their use in commercial application is growing. Drones are considered as a first choice for applications such as search-and-rescue, surveillance, and wildlife conservation. These applications usually require real-time video analytics which is a major challenge. Although the use of edge computing in drone-based video applications can bring significant improvement in latency and offloading intensive computation tasks, live video analytics is still the area which requires further research and state-of-the-art solutions. In particular these issues need further attention. (i) Drone's mobility may need computation migration from one edge server to another on the fly. (ii) The wireless links from drone to the edge server rapidly vary and the resource allocation shall be mandatory considering a joint communication and computation mechanism. (iii) Drones have limited onboard energy and only minimal computation to be performed locally to reduce energy consumption of drone. (iv) Drones that are receiving poor coverage by all edge servers may rely on D2D offloading to forward computation or caching access to other drones in the coverage. However, the effectiveness of such D2D cooperation is highly dependent upon the aerial links quality and stability. MEC-assisted drones \cite{wang2019edge} and drone-drone cooperation in video services involve large-sized content sharing and intensive offloading, which poses severe challenges as compared to normal drone cooperation. 

\subsection{MEC in VR, AR, MR and $360^{\circ}$ Videos Streaming}
Virtual Reality, Augmented Reality and Mixed Reality have been briefly discussed earlier in the paper. Recently $360^{\circ}$ videos have been introduced, as a part of VR. $360^{\circ}$ videos are large sized videos typically about 5x larger than that of conventional videos. \emna{Furthermore, the $360^{\circ}$ videos require lower delivery latency and higher bandwidth compared to traditional video streaming. More importantly, the Quality of Experience (QoE) of viewers is highly sensitive to the dynamics of the network environment and it is extremely intolerant to delay variance and image freezing.  Moreover, because of the huge increase of the number of VR devices which are estimated to reach 50 million in 2021 \cite{360}, the backhaul network becomes a bottleneck.  As stated previously,  the cloud wisdom is no longer sustainable for such real-time applications with stringent requirements. Therefore, MEC edge servers are introduced as an alternative to the cloud computing offering computation and caching capabilities in the vicinity of end-users. However, the huge processing requirements of $360^{\circ}$ videos could quickly exhaust the available  resources of the MEC Servers. Hence, efficient algorithms need to be designed for the given computational resources. Second, caching multiple views of video incurs high overhead in storage. Even if hard disks are not expensive nowadays, storing all of these files is neither economical nor feasible.}

\eb{
\subsection{Secure live video streaming using Blockchain}
To address the challenges of lives video streaming, the collaborative MEC servers process raw streams locally to serve viewers with minimal delays. However, the wireless data transmission incurs concerns in data security and privacy because it exposes vulnerabilities to potential attackers to perform malicious operations, such as eavesdropping private video streams (e.g. non-paying subscribers). To prevent the attackers from hacking the data, it is necessary to secure the communication channel between different edge nodes through encryption. However, due to the limited computational resources available at MEC servers and the time sensitivity of live streaming, the encryption will be relatively weak. Since the live streaming system is composed of a distributed network environment with a huge number of crowdsourcers with large heterogeneity, a dynamic, flexible, scalable and lightweight security mechanism is required. Additionally, the crowdsourcers (subscribers) are geographically distributed across an untrusted edge network. Hence, it is not
suitable to implement security system on a centralized device, which may suffer from performance bottleneck. Blockchain \cite{chain} is a distributed database that contains chronologically connected data blocks. Each block is an individual component that includes information linked to a specific transaction. We envision that blockchain will be a strong fit to provide a suitable solution for secure live video sharing because of its immutability and decentralization features, which are perfectly consistent with our context. New requirements need to be handled in this task including, addressing the trade-off between increasing security level and latency. Public blockchain is slower than traditional database, since it coordinates the blocks of multiple unaffiliated participants, which contradicts the immediate responsiveness of live streaming. However, since we deal with already
subscribed users of paid channels, blockchain solution can be feasible. A future direction could be the design of a complete architecture for blockchain-enabled authentication scheme for live videos systems in MEC networks and a comprehensive evaluation to validate the feasibility of the proposed authentication on the live streaming platforms.
}


\section{Conclusions}
\label{sec:conclusions}
This paper surveys the research efforts on mobile edge computing for video streaming services. The paper presented a brief overview of mobile edge computing architecture and the various benefits it brings to users and network/content providers. Different schemes for edge-based content caching and processing are discussed and categorized. The potential benefits of mobile edge computing for video streaming services are presented. The state-of-the-art in MEC for video streaming is summarized. In addition to the MEC applications for video streaming services, D2D cooperation among end users to improve the edge computing performance is presented and research efforts in D2D-streaming are discussed. Machine learning techniques have been rigorously used in MEC systems and a significant amount of work has been done in this area. The paper summarised the use of ML/DL in MEC systems for video streaming applications. Lastly, some useful insights and future research directions are provided for prospective researchers.

\section*{Acknowledgement}
This paper was made possible by Qatar University Internal Grant No. IRCC-2020-001. The statements made herein are solely the responsibility of the authors.

\bibliography{biblio,biblio_survey,biblio_ml,biblio_stream, biblio_mec_stream, biblio_d2d, biblio_new}
\bibliographystyle{ieeetr}


\begin{IEEEbiography}[{\includegraphics[width=1in,height=1.25in,clip,keepaspectratio]{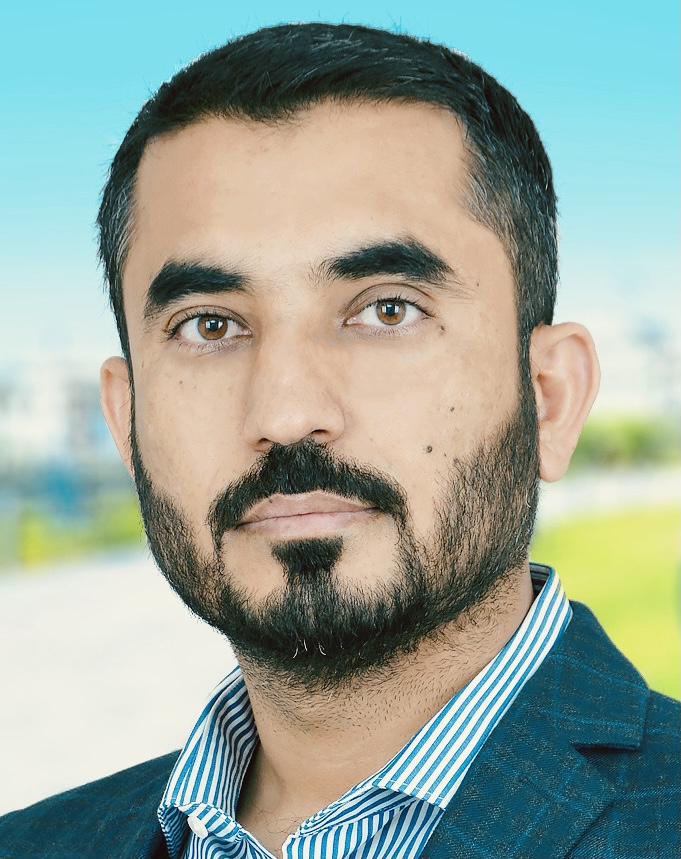}}]{Muhammad Asif Khan} (SM'20) received the B.Sc. degree in Telecommunication Engineering from University of Engineering and Technology Peshawar, Pakistan (2009) and M.Sc. in Telecommunication Engineering from University of Engineering and Technology Taxila, Pakistan (2013). He received the Ph.D. degree in Electrical Engineering from Qatar University in 2020. He was a Researcher Assistant at Qatar University (2014-2015) and at Qatar Mobility Innovation Center (2016-2017). He is currently working as a postdoctoral researcher at Qatar University. His current research interests include wireless communication, mobile edge computing, machine learning and distributed optimization. He is a Senior Member of IEEE and a Member of IET. For more detailed information, please visit his homepage: \url{http://www.asifk.me}.
\end{IEEEbiography}

\vskip -1\baselineskip plus -1fil

\begin{IEEEbiography}[{\includegraphics[width=1in,height=1.25in,clip,keepaspectratio]{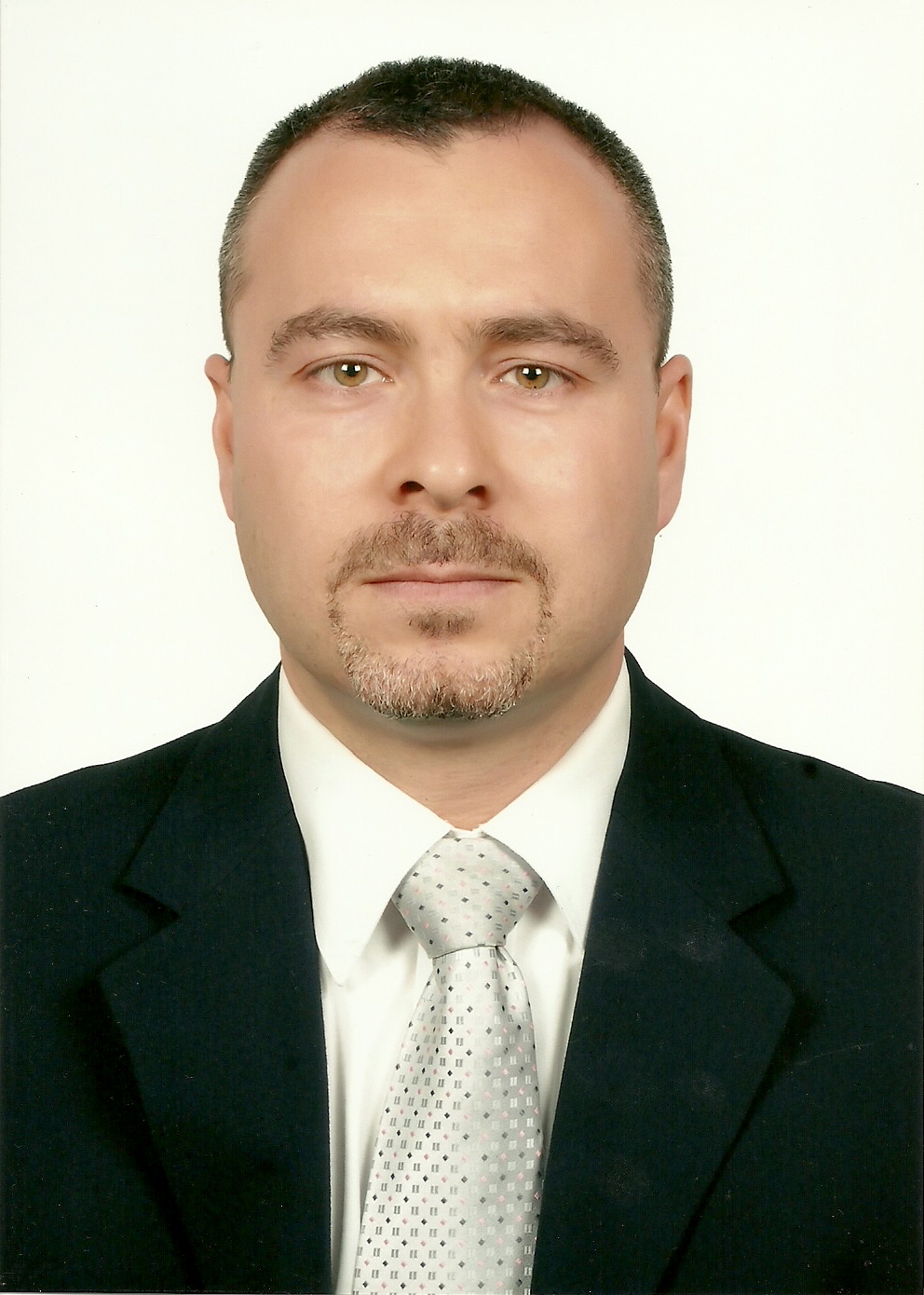}}]{Prof. Ridha Hamila} received the MSc, LicTech with distinction, and PhD degrees from Tampere University, Tampere, Finland, in 1996, 1999, and 2002, respectively. Dr. Hamila currently a Full Professor at the Department of Electrical Engineering, Qatar University, Qatar. From 1994 to 2002 he held various research and teaching positions at TUT within the Department of Information Technology, Finland. From 2002 to 2003 he was a System Specialist at Nokia research Center and Nokia Networks, Helsinki. From 2004 to 2009 he was with Emirates Telecommunications Corporation, UAE. Also, from 2004 to 2013 he was adjunct Professor at the Department of Communications Engineering, TUT. His current research interests include mobile and broadband wireless communication systems, Mobile Edge Computing, Internet of Everything, and Machine Learning. In these areas, he has published over 200 journal and conference papers most of them in the peered reviewed IEEE publications, filed seven US patents, and wrote numerous confidential industrial research reports.  Dr. Hamila has been involved in several past and current industrial projects, Ooreedo, Qatar National Research Fund, Finnish Academy projects, EU research and education programs. He supervised a large number of under/graduate students and postdoctoral fellows. He organized many international workshops and conferences. He is a Senior Member of IEEE.
\end{IEEEbiography}

\vskip -1\baselineskip plus -1fil

\begin{IEEEbiography}[{\includegraphics[width=1in,height=1.25in,clip,keepaspectratio]{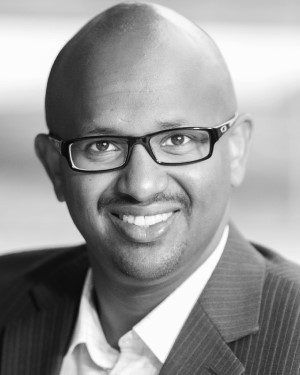}}]{Aiman Erbad} (Senior Member, IEEE) received the MSc degree from the University of Essex (2005), and the Ph.D degree from the University of British Columbia (2012). He is currently an Associate Professor with the College of Science and Engineering, Hamad Bin Khalifa University (HBKU). His research interests include cloud computing, edge computing, the IoT, private and secure networks, and multimedia systems. He received the Platinum award from H. H. Emir Sheikh Tamim bin Hamad Al Thani at the Education Excellence Day 2013 (Ph.D. category). He also received the 2020 Best Research Paper Award from Computer Communications, the IWCMC 2019 Best Paper Award, and the IEEE CCWC 2017 Best Paper Award. His research interest spans cloud computing, edge computing, IoT, distributed AI algorithms, and private / secure networks. Aiman is currently an editor in the International Journal of Sensor Networks (IJSNet), an editor in KSII Transactions on Internet and Information Systems, and served as a guest editor in IEEE Networks.
\end{IEEEbiography}

\vskip -1\baselineskip plus -1fil

\begin{IEEEbiography}[{\includegraphics[width=1in,height=1.25in,clip,keepaspectratio]{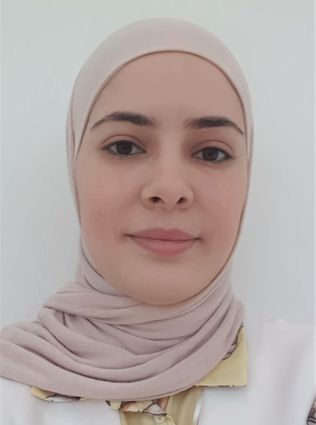}}]{Emna Baccour} received the Ph.D. degree in computer Science from the University of Burgundy, France, in 2017. She was a postdoctoral fellow at Qatar University on a project covering the interconnection networks for massive data centers and then on a project covering video caching and processing in mobile edge computing networks. She currently holds a postdoctoral position at Hamad Ben Khalifa University. Her research interests include data center networks, cloud computing, green computing and software defined networks as well as distributed systems. She is also interested in edge networks, mobile edge caching and computing, and IoT systems.
\end{IEEEbiography}

\vskip -1\baselineskip plus -1fil

\begin{IEEEbiography}[{\includegraphics[width=1in,height=1.25in,clip,keepaspectratio]{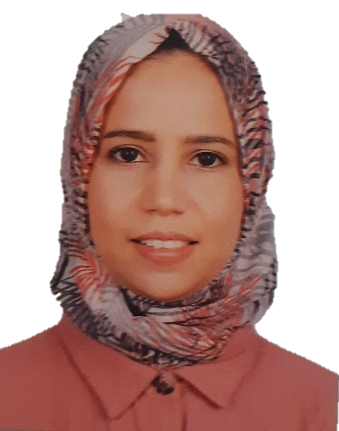}}]{Zina Chkirbene} received her Ph.D. in Computer science in 2017 from the University of Burgundy Dijon, France. She received her Bachelor's degree in Computer Science Networks and Telecommunications from the National Institute of Applied Science and Technology, in 2011, and received her master's degree in Electronic Systems and Communication Networks from Polytechnic School, in 2012.   She was a Research Assistant at Qatar University on a project covering the interconnection networks for massive data centers. She currently holds a postdoctoral position at Qatar University. Her research interests include data center networks, edge computing, green computing, and machine learning as well as deep reinforcement learning techniques.
\end{IEEEbiography}

\vskip -1\baselineskip plus -1fil

\begin{IEEEbiography}[{\includegraphics[width=1in,height=1.25in,clip,keepaspectratio]{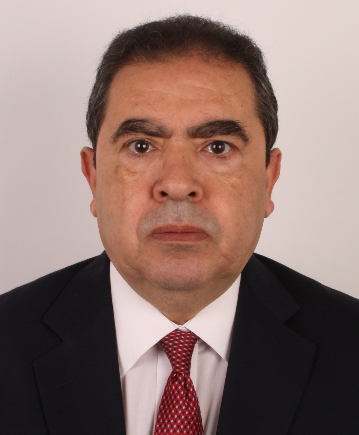}}]{Prof. Mounir Hamdi} is the founding Dean of the College of Science and Engineering at Hamad Bin Khalifa University (HBKU). He is an IEEE Fellow for contributions to design and analysis of high-speed packet switching. As founding Dean of the College of Science and Engineering, Dr. Hamdi led the foundation of 15 graduate programs and 1 undergraduate program and all the associated research labs and activities. Before joining HBKU, he was Chair Professor at the Hong Kong University of Science and Technology (HKUST), and the Head of the Department of Computer Science and Engineering. In 1999 to 2000 he held visiting professor positions at Stanford University, USA, and the Swiss Federal Institute of Technology, Lausanne, Switzerland. His general area of research is in high-speed wired/wireless networking in which he has published more than 400 research publications, received numerous research grants, and graduated more 50 MS/PhD students. Prof. Hamdi is/was on the Editorial Board of various prestigious journals and magazines including IEEE Transactions on Communications, IEEE Communication Magazine, Computer Networks, Wireless Communications and Mobile Computing, and Parallel Computing. In addition to his commitment to research and professional service, he is also frequently involved in higher education quality assurance activities as well as engineering programs accreditation all over the world.
\end{IEEEbiography}

\vskip -1\baselineskip plus -1fil

\begin{IEEEbiography}[{\includegraphics[width=1in,height=1.25in,clip,keepaspectratio]{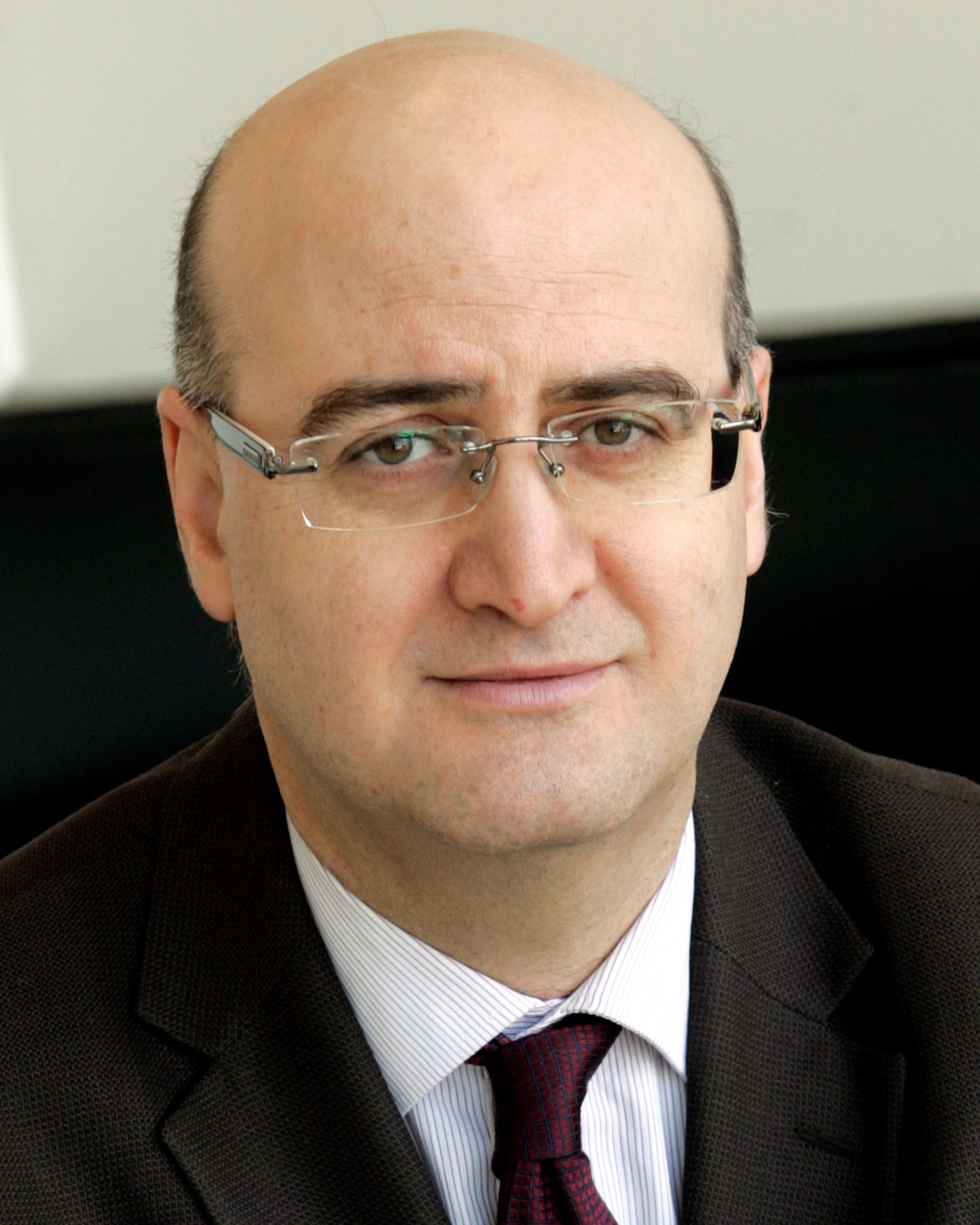}}]{Prof. Moncef Gabbouj} (F’11) is a well-established world expert in the field of image processing, and held the prestigious post of Academy of Finland Professor during 2011-2015. He has been leading the Multimedia Research Group for nearly 25 years and managed successfully a large number of projects in excess of 18M Euro. He has supervised 45 PhD theses and over 50 MSc theses. He is the author of several books and over 700 papers. His research interests include Big Data analytics, multimedia content-based analysis, indexing and retrieval, artificial intelligence, machine learning, pattern recognition, nonlinear signal and image processing and analysis, voice conversion, and video processing and coding. Dr. Gabbouj is a Fellow of the IEEE and member of the Academia Europaea and the Finnish Academy of Science and Letters. He is the past Chairman of the IEEE CAS TC on DSP and committee member of the IEEE Fourier Award for Signal Processing. He served as associate editor and guest editor of many IEEE, and international journals.
\end{IEEEbiography}

\balance

\end{document}

%% file: tables/surveys2.tex
   \begin{tabular}{|l|m{9cm}|c|c|c|c|c|}
   \hline
    \textbf{Ref}   &\textbf{One-sentence summary}  & \multicolumn{5}{c|}{\textbf{Scope of the Publication}} \\[0.8em] \cline{3-7} 
          &                      &\rot{Architecture} &\rot{Caching}  &\rot{Offloading}  &\rot{ML/DL}  &\rot{Video streaming} \\[0.8em] \hline

    Wang \etal \cite{wang_2020} &Deep learning in edge computing &\xmark &\xmark &\xmark &\checkmark &\xmark \\[0.5em] \hline
    
    Deng \etal \cite{deng_2020} &Artificial Intelligence (AI) in edge computing &\xmark &\xmark &\xmark &\checkmark &\xmark \\[0.5em] \hline
    
    Carvalho \etal \cite{carvalho_2020} &Using AI for computation offloading in edge computing for &\checkmark &\xmark &\checkmark &\checkmark &\xmark \\[0.5em] \hline
    
    Cao \etal \cite{cao_2020} &Overview of edge computing research &\checkmark &\xmark &\xmark &\xmark &\xmark \\[0.5em] \hline
    
    Lin \etal \cite{lin_2020} &A survey of computation offloading techniques in edge computing &\checkmark &\xmark &\checkmark &\checkmark &\xmark \\[0.5em] \hline
    
    Shakarami \etal \cite{shakarami_2020} &A survey of computation offloading techniques in edge and cloud computing &\checkmark &\xmark &\checkmark &\xmark &\xmark \\[0.5em] \hline
    
    
    Xiao \etal \cite{xiao_2019} &A survey on security challenges in edge computing &\checkmark &\xmark &\xmark &\xmark &\xmark \\[0.5em] \hline 
    
    Yang \etal \cite{yang_2019} &A survey on integration of bloackchains and edge computing &\checkmark &\xmark &\xmark &\xmark &\xmark \\[0.5em] \hline
    
    Liu \etal \cite{liu_2019} &A survey on secured edge-based data analytics &\checkmark &\xmark &\xmark &\xmark &\xmark \\[0.5em] \hline
    
    Jiang \etal \cite{jiang_2019} &A survey of computation offloading in edge computing &\checkmark &\xmark &\checkmark &\xmark &\xmark \\[0.5em] \hline
    
    Lin \etal \cite{lin_2019} &A survey of computation offloading in edge computing &\checkmark &\xmark &\checkmark &\xmark &\xmark \\[0.5em] \hline
    
    Donno \etal \cite{de_2019} &A tutorial on modern computing paradigms &\checkmark &\xmark &\xmark &\xmark &\xmark \\[0.5em] \hline
    
    Khan \etal \cite{khan_2019} &A survey on edge computing applications  &\checkmark &\xmark &\checkmark &\xmark &\xmark \\[0.5em] \hline
    
    Hassan \etal \cite{hassan_2019} &A review of edge computing in 5G networks &\xmark &\xmark &\checkmark &\xmark &\xmark \\[0.5em] \hline
    
    Chen \etal \cite{chen_2019} &A review of deep learning applications in edge computing &\xmark &\xmark &\checkmark &\checkmark &\xmark \\[0.5em] \hline
    
    Zhang \etal \cite{zhang_2018} &A survey on security and privacy in edge computing and &\checkmark &\xmark &\xmark &\checkmark &\xmark \\[0.5em] \hline
    
    Marjanovic \etal \cite{marjanovic_2018} &A tutorial on edge computing for crowdsensing &\checkmark &\xmark &\xmark &\xmark &\xmark \\[0.5em] \hline
    
    Porambage \etal \cite{porambage_2018} &A survey on application of edge computing in Internet of Things (IoT) &\checkmark &\xmark &\xmark &\xmark &\xmark \\[0.5em] \hline 
    
    Abbas \etal \cite{abbas_2018} &A tutorial and survey on mobile edge computing &\checkmark &\xmark &\xmark &\xmark &\xmark \\[0.5em] \hline
    
    Wang \etal \cite{wang_2017} &A survey on edge computing services &\checkmark &\checkmark &\checkmark &\xmark &\xmark \\ \hline
    
    Baktir \etal \cite{baktir_2017} &A survey on software-defined networking in edge computing &\checkmark &\xmark &\xmark &\xmark &\xmark \\[0.5em] \hline
    
    Ahmed \etal \cite{ahmed_2017} &A survey of the state-of-the-art in mobile edge computing &\xmark &\checkmark &\xmark &\xmark &\xmark \\[0.5em] \hline
    
    Taleb \etal \cite{taleb_2017} &A survey of edge computing in 5G networks &\checkmark &\checkmark &\checkmark &\xmark &\xmark \\[0.5em] \hline
    
    Mao \etal \cite{mao_2017} &A review of the state-of-the-art in mobile edge computing &\checkmark &\xmark &\checkmark &\xmark &\xmark \\[0.5em] \hline
    
    Mao \etal \cite{mao_2017_2} &A survey of communication techniques in mobile edge computing &\xmark &\xmark &\xmark &\checkmark &\xmark \\[0.5em] \hline
    
    Shirazi \etal \cite{shirazi_2017} &A tutorial on security in mobile edge and fog computing &\checkmark &\xmark &\xmark &\xmark &\xmark \\[0.5em] \hline
    
    Yu \etal \cite{yu_2017} &A survey and tutorial on edge computing in IoT &\checkmark &\xmark &\checkmark &\xmark &\xmark \\[0.5em] \hline
    
    Ahmed \etal \cite{ahmed_2016} &A survey of mobile edge computing systems &\checkmark &\xmark &\xmark &\xmark &\xmark \\[0.5em] \hline  
    
    Shi \etal \cite{shi_2016} &A futuristic overview of mobile edge computing &\checkmark &\xmark &\checkmark &\xmark &\xmark \\[0.5em] \hline
    
    
    \khan{Jiang \etal} \cite{jiang2021} &\khan{MEC in Video Streaming} &\checkmark &$\square$ &\checkmark &\xmark &\checkmark \\[0.5em] \hline
    
    \khan{Kanai \etal} \cite{Kanai2018} &\khan{MEC for multimedia applications} &\xmark &$\square$ &$\square$ &\xmark &\xmark \\[0.5em] \hline
    
    \khan{Zhang \etal} \cite{zhang2019} &\khan{MEC for video analytics in public safety} &$\square$ &\xmark &\xmark &\xmark &$\square$ \\[0.5em] \hline

    \rowcolor{red!10}
    \khan{This survey}& \khan{A survey of MEC in video streaming covering the MEC architecture, video streaming applications and novel services, and traditional and state-of-the-art techniques for edge-based caching and processing. It also covers the use of ML and pervasive computing including D2D for the next generation intelligent edge services.} &\checkmark &\checkmark &\checkmark &\checkmark &\checkmark \\[0.8em] \hline
    
    \end{tabular}

    
    
    
    

    

%% file: tables/abbrev.tex
\begin{tabular}{| l | l |} \hline
\textbf{Acronym} &\textbf{Definition} \\[0.5em]
\hline \hline

3D &Three Dimensional \\\hline
3GPP &Third Generation Partnership Project \\\hline
4K &Horizontal video resolution of 4000 pixels \\\hline
5G &Fifth-Generation \\\hline
ABR &Adaprive Bit Rate \\\hline
AI &Artificial Intelligence Internet \\\hline
AR &Augmented Reality \\\hline
BS &Base Station \\\hline
CC &Cloud Computing \\\hline
CDN &Content Delivery Network \\\hline
CMAF &Common Media Application Format \\\hline
CNN &Convolutional Neural Network Recurrent \\\hline
D2D &Device to Device \\\hline
DL &Deep Learning \\\hline
DNN &Deep Neural Network \\\hline
\asif{DOS} &\asif{Denial Of Service} \\ \hline
DQN &Deep Q-learning Network \\\hline
DRL &Deep Reinforcement Learning \\\hline
EC &Edge Computing \\\hline
ETSI &European Telecommunication Standards Institute \\\hline
\asif{FIFO} &\asif{First In First Out} \\ \hline
FOV &Field of View \\\hline
HD &High Definition \\\hline
HDS &HTTP Dynamic Streaming \\\hline
HLS &HTTP Live Streaming \\\hline
\asif{HTTP} &\asif{Hypertext Transfer Protocol} \\\hline
ILP &Integer Linear Programming \\\hline
INLP &Integer Non-Linear Programming \\\hline
IoT &Internet of Things \\\hline
ISG &Industry Specification Group \\\hline
ISG &Industry Specific Groups \\\hline
LFU &Least Frequently Used \\ \hline
LLR &Least Likely Requested \\\hline
LRU &Least Recently Used \\\hline
LRFU &Least Recently Frequency Used \\ \hline
LSTM &Long-short Term Memory model \\\hline
MDP &Markov Decision Problem\\\hline
MEC &Mobile Edge Computing \\\hline
MEC &Mobile (or Multi-Access) Edge Computing \\\hline
\asif{MILP} &\asif{Mixed Integer Linear Programming} \\ \hline
ML &Machine Learning \\\hline
MLR &Most Likely Requested \\\hline
DASH &Dynamic Adaptive Streaming over HTTP \\\hline
MPV &Most Popular Videos \\\hline
\asif{MPEG} &\asif{Motion Picture Expert Group} \\ \hline
MSS &Microsoft Smooth Streaming \\\hline
NFV &Network Function Virtualization \\\hline
\asif{PcP} &\asif{Proactive Cache Policy} \\ \hline
P-UPP &Proactive User Preference Profile \\\hline
\asif{PoP} &\asif{Point of Presence} \\ \hline
QoE &Quality of Experience \\\hline
QoS &Quality of Service \\\hline
R-UPP &Reactive User Preference Profile \\\hline
RAN &Radio Access Network \\\hline
RFC &Request For Comments \\\hline
RL &Reinforcement Learning \\\hline
RNN &Recurrent Neural Network \\\hline
RTMP &Real-Time Messaging Protocol \\\hline
RTP &Real-Time Transport Protocol \\\hline
RTSP &Real-Time Streaming Protocol \\\hline
\asif{SIP} &\asif{Session Initiation Protocol} \\ \hline
SRT &Secure Reliable Transport \\\hline
TDMA &Time Devision Multiple Access \\\hline
UA &User Attrition \\\hline
UAV &Unmanned Aerial Vehicle \\\hline
UPP &User Preference Profile \\\hline
URRLC &Ultra-reliable Low-latency Communications \\\hline
V2I &Vehicle to Infrastructure \\\hline
VoD &Video On Demand \\\hline
VR &Virtual Reality \\\hline
WebRTC &Web Real-Time Communications \\\hline
\end{tabular}

%% file: tables/stream_proto.tex
\begin{tabular}{|c|c|c|c|c|}
\hline 
Protocol &Developer &Year &Latency(s)  &ABR \\[0.5em] \hline \hline

RTMP               &Macromedia  &2012    &Low &\checkmark\\

RTSP/RTP           &IETF    &1998     &Ultra low   &\xmark  \\

HLS                &Apple       &2009    &Reduced  &\checkmark \\ 

LL-HLS              &Apple      &2009    &Low &\checkmark \\

MPEG-DASH          &MPEG        &2011     &High &\checkmark \\ 

CMAF               &MPEG        &2018    &Low  &\checkmark \\ 

MSS                &Microsoft   &2008    &Reduced  &\checkmark \\

HDS v3             &Adobe       &2013    &Reduced  &\checkmark  \\

SRT                &Haivision   &2013    &Low  &\checkmark  \\

SIP/RTP            &IETF        &1999    &Reduced  &\checkmark  \\

WebRTC             &IETF        &2011    & Ultra low &\checkmark \\

\hline
\end{tabular}

%% file: tables/resource_alloc1.tex
\begin{tabular}{| m{2cm} | m{0.5cm} | m{8cm} | m{2.5cm} | m{2.5cm} |}
\hline
\textbf{Category}  &\textbf{Ref} &\textbf{Description} &\khan{\textbf{Evaluation Metrics}} & \khan{\textbf{Method/Algorithm}}\\[1em] \hline \hline

& \cite{barbarossa_2014} & Transmission energy minimization under computation deadline. &\khan{Delay, Energy}  &\khan{Convex programming}\\
Binary Offloading & \cite{zhang_2013} & Minimize energy consumption with a soft real-time requirement. &\khan{Delay, Energy}  &\khan{Convex Programming}\\
& \cite{Hao2019} & A binary offloading model to maximize profit to video service provider. &\khan{N/A} &\khan{MAB}
\\[.5em] \hline

& \cite{wei_2018} & Schedule offloading tasks from multiple mobile nodes to a single MEC server in order to maximize energy savings of all mobile nodes. &\khan{Energy}  &\khan{SMSEF}\\ 
Partial Offloading  & \cite{wang_2016} & Task-input data is divided for local and remote execution. &\khan{Energy, Latency}  &\khan{uni-variate search}\\ 
& \cite{jia_2014} & Load balancing between mobile and edge servers to minimize latency. &\khan{Latency}  &\khan{Heuristic}
\\[.5em] \hline

& \cite{yang_2018} & Reinforcement learning to proactive allocation resources to multiple. &\khan{E2E Reliability}  &\khan{Q-learning}\\
Stochastic Offloading & \cite{haouari_2019_2} & Transcoding resource allocation using past video requests and then using time series forecasting to predict the resources needed for future requests using a heuristic. &\khan{QoE, Network cost}  &\khan{LSTM, GRU, CNN, MLP, XGboost}
\\[.5em] \hline

& \cite{mehrabi_2019} & A scheduling algorithm that solves client to edge mapping (load balancing) and per client bit rate selection problems. &\khan{Throughput, Buffer delay, Fairness}  &\khan{Heuristic} \\
Server Scheduling & \cite{molina_2014} & Scheduling uplink and downlink transmissions using queuing theory. &\khan{Energy}  &\khan{Nested interval algorithm}\\  
& \cite{guo_2016} & Tasks Offloading selection using clock frequency and transmission power allocation. &\khan{Energy, Delay}  &\khan{DVFS}
\\[.5em] \hline

Joint Radio and Processing
& \cite{ren_2019} & Jointly optimize communication and computational resources with tasks splitting between edge and cloud. &\khan{Latency}  &\khan{Convex optimization}\\ 
& \cite{barbarossa_2013} & Joint allocation of computation and radio resources. &\khan{Power consumption}  &\khan{Convex optimization}
\\[.5em] \hline

& \cite{tran_2017} & Jointly optimize caching and transcoding resources to minimize the backhaul network cost. &\khan{Network cost}  &\khan{ILP} \\
Joint Caching and Processing &\cite{Ge2017} & Collaborative offloading and caching scheme using Lyapunov optimization to minimize the overall latency of all mobile devices. &\khan{Throughput, Playout delay, Buffer length, Rebuffering duration}  &\khan{Theoretical architecture}
\\[.5em] \hline

Server Selection
& \cite{niu_2019} & Balance work load among edge nodes by considering edge computing capability, existing work load and distance among edge server and user. &\khan{Service delay}  &\khan{PSO}\\ 
& \cite{haouari_2019} & Proactively allocate resources by predicting the number of viewers in a cloud site and then serving viewers by their closest server. &\khan{Viewers' QoE, Network cost}  &\khan{MLP, DT, RF}
\\[.5em] \hline

& \cite{bilal_2019} & edge servers in a sing-hop clusters cooperate to cache and transcode contents. &\khan{Network cost, delay, cache hit ratio}  &\khan{ILP}\\  
Server Cooperation & \cite{ren_2019} & Jointly optimize communication and computational resources with tasks splitting between edge and cloud. &\khan{Delay}  &\khan{Convex optimization}\\  
& \cite{zhao_2015} & Edge/cloud selection to meet deadline requirements. &\khan{Task completion}  &\khan{Heuristic}
\\[.5em] \hline

& \cite{wang_2014} &Computation migration based on distance threshold between user and two servers using MDP. &\khan{MDP}  &\khan{Delay}\\ 
Computation Migration & \cite{urgaonkar_2015} & Jointly optimize computation scheduling and service migration to minimize energy transmission and reconfiguration cost. &\khan{MDP}  &\khan{Application queue length} \\  
& \cite{chen_2016} & Compute locally or offload to remote server such that total energy consumption and latency is minimized. &\khan{Energy, Delay}  &\khan{MIP}
\\[1.5em] \hline
\end{tabular}

%% file: tables/mec_apps_stream.tex
\begin{tabular}{| m{2cm}  p{5cm}|}
    \hline
    \rowcolor[HTML]{32995F}
    \multicolumn{2}{|c|}{\textcolor{white}{\textbf{Content Searching}}}\\[0.5em]
    \hline
    Faster search 
    &Faster content search with user's search history cached and updated over time.\\[1.5em]
    
    Personalised search
    &Personalized search based on user's location (e.g. supermarket, office, stadium) \\[1em]
    \hline
\end{tabular}

\bigskip
\begin{tabular}{| m{2cm}  p{5cm}|}
    \hline
    \rowcolor[HTML]{15AEBF}
    \multicolumn{2}{|c|}{\textcolor{white}{\textbf{Content Suggestions}}}\\[0.5em]
    \hline
    Use case 1  
    &Personalized content suggestion based on user preferences.\\[1.5em]
    
    Use case 2
    &Improved content suggestions over time when user preferences changes. \\[1.5em]
    
    Use case 3
    &Content suggestions based on local/regional events. \\[1em]
    \hline
\end{tabular}

\bigskip
\begin{tabular}{| m{2cm}  p{5cm}|}
    \hline
    \rowcolor[HTML]{736CA8}
    \multicolumn{2}{|c|}{\textcolor{white}{\textbf{Targeted Advertisements}}}\\[0.5em]
    \hline
    Use case 1 
    &Highly targeted advertisements based on user's interests.\\[1.5em]
    Use case 2
    &Targeted advertisements based on user's previously visited shopping locations with timing information.\\[1em]
    \hline
\end{tabular}

\bigskip
\begin{tabular}{| m{2cm}  p{5cm}|}
    \hline
    \rowcolor[HTML]{CC0066}
    \multicolumn{2}{|c|}{\textcolor{white}{\textbf{Interactive Videos}}}\\[0.5em]
    \hline
    \rowcolors{white}{gray}
    LLive Sports 
    &Display Live statistics of a match\\[1em]
    
    Filmography
    & Pulling filmography of an actor appearing in a scene \\[1.5em]

    Retail/E-commerce
    & Allowing users to click on objects inside video to see details or buy. \\[1.5em]
    
    Education/Trainings
    & Allow users to ask questions, submit answers, fill surveys\\
    
    \hline
\end{tabular}

\bigskip
\begin{tabular}{| m{2cm} p{5cm}|}
    \hline
    \rowcolor[HTML]{777777}
    \multicolumn{2}{|c|}{\textcolor{white}{\textbf{Video Analytics}}}\\[0.5em]
    \hline
    Surveillance 
    &object detection, motion tracking, facial recognition, gesture recognition , activity recognition, head counting\\[1.5em]
    
    Content detection
    &detecting banned videos, parental control, illegal videos\\[1.5em]
    
    QoE Measurement
    &KPI calculation for encrypted streaming videos\\[1.5em]
    
    Realtime Assistance
    &Detecting obstacles using videos captured by stereoscopic camera\\[1em]
    \hline
\end{tabular}

\bigskip
\begin{tabular}{| m{2cm} p{5cm}|}
    \hline
    \rowcolor[HTML]{E85642}
    \multicolumn{2}{|c|}{\textcolor{white}{\textbf{VR and AR}}}\\[0.5em]
    \hline
    Retail 
    &Try items using AR/VR reality without physically trying things\\[1.5em]
    
    Gaming
    &Pokémon GO is a popular example.\\[1.5em]
    
    Education/Trainings
    & Wearable cognitive assistance \\[1.5em]
    
    Tourism
    & Try items using AR/VR reality without physical try things \\[1.5em]
    
    Healthcare
    & AR enabled minimal invasive surgeries \\[1.5em]
    
    Industry/safety
    &Building Digital maps and digital twin \\[1em]
    \hline
\end{tabular}

%% file: tables/mec_stream_soa.tex
\begin{tabular}{|m{2cm}|>{\centering\arraybackslash}m{0.5cm}|m{13cm}|c|c|c|c|c|c|c|}
\hline 
\multirow{2}{*}{Category} &\multirow{2}{*}{Ref} &\multirow{2}{*}{Description of Contribution}
& \multicolumn{7}{c|}{\khan{Evaluation Metrics}} \\[0.8em] \cline{4-10} 
          & & &\rot{\khan{Delay}}
          &\rot{\khan{Energy}}
          &\rot{\khan{Hit ratio}}
          &\rot{\khan{Network cost}}
          &\rot{\khan{Capacity}}
          &\rot{\khan{Revenue}}
          &\rot{\khan{Other}} \\[0.8em] \hline \hline


\multirow{9}{2cm}{Caching and Proactive Content Fetching} 
&\cite{Krishnappa2011} &Fetch Most Popular Videos (MPV) using nation-wide popularity. &\khan{\xmark} &\khan{\xmark} &\khan{\xmark} &\khan{\xmark} &\khan{\checkmark} &\khan{\xmark} &\khan{\xmark} \\ \cline{2-10}
&\cite{gharaibeh_2016}  &Proactively fetch contents to reduce the sum of storage cost and user attrition cost. &\khan{\xmark} &\khan{\xmark} &\khan{\xmark} &\khan{\checkmark} &\khan{\xmark} &\khan{\xmark} &\khan{\xmark} \\\cline{2-10}
&\cite{Tran2020} & Caching based on popularity and retention rate of video streams to maximize their video bitrate. &\khan{\xmark} &\khan{\xmark} &\khan{\xmark} &\khan{\xmark} &\khan{\checkmark} &\khan{\xmark} &\khan{\xmark} \\\cline{2-10}
&\cite{Zeng2019} & Smart caching based on user behavior to maximize the hit rate of contents within limited time. &\khan{\xmark} &\khan{\xmark} &\khan{\checkmark} &\khan{\xmark} &\khan{\xmark} &\khan{\xmark} &\khan{\xmark} \\\cline{2-10}
&\cite{Su2017} & Cache layered videos for multiple user groups. Users in same group share caching cost. &\khan{\checkmark} &\khan{\xmark} &\khan{\checkmark} &\khan{\xmark} &\khan{\xmark} &\khan{\xmark} &\khan{\xmark} \\\cline{2-10}
&\cite{Yang2019} & Prefetch popular videos from YouTube using ML-based prediction. &\khan{\xmark} &\khan{\xmark} &\khan{\checkmark} &\khan{\xmark} &\khan{\checkmark} &\khan{\xmark} &\khan{\checkmark} \\\cline{2-10}
&\cite{Kumar2020} & Cache videos that serves the maximum users by estimating the requested qualities using RAN information. &\khan{\checkmark} &\khan{\xmark} &\khan{\checkmark} &\khan{\xmark} &\khan{\checkmark} &\khan{\xmark} &\khan{\xmark} \\\cline{2-10}
& \khan{\cite{yang2021_2}} & \khan{Edge caching strategy for vehicular networks in which video content is stored on the RSU and fetched by users.} &\khan{\checkmark} &\khan{\xmark} &\khan{\checkmark} &\khan{\xmark} &\khan{\xmark} &\khan{\xmark} &\khan{\xmark} \\ [0.0em]  \hline \hline

\multirow{18}{2cm}{Cooperative Caching}
&\cite{bilal_2019} & Prefetch highest bitrate version of the video from CDN server using X2 interface. &\khan{\checkmark} &\khan{\xmark} &\khan{\checkmark} &\khan{\checkmark} &\khan{\xmark} &\khan{\xmark} &\khan{\xmark} \\\cline{2-10}
&\cite{Vigneri2019} & Prefetch video chunks into playout buffer from encountered vehicle caches/ stream from the cellular. &\khan{\xmark} &\khan{\xmark} &\khan{\xmark} &\khan{\xmark} &\khan{\xmark} &\khan{\xmark} &\khan{\checkmark} \\\cline{2-10}
&\cite{Qu2020} & Solve the multiple bitrate video caching problem using polynomial complexity algorithm. &\khan{\xmark} &\khan{\xmark} &\khan{\xmark} &\khan{\xmark} &\khan{\xmark} &\khan{\xmark} &\khan{\checkmark} \\\cline{2-10}
&\cite{Zhang2020a} & Operators advertise contents to attract a mobile user to request cached contents instead of non-cached contents. &\khan{\xmark} &\khan{\xmark} &\khan{\xmark} &\khan{\xmark} &\khan{\xmark} &\khan{\checkmark} &\khan{\xmark} \\\cline{2-10}
&\cite{Cheng2020} & Context-aware adaptive caching through network virtualization at the edge. &\khan{\xmark} &\khan{\checkmark} &\khan{\xmark} &\khan{\xmark} &\khan{\xmark} &\khan{\xmark} &\khan{\xmark} \\\cline{2-10}
&\cite{Li2020a} & Jointly optimize caching of bitrate-aware files and the scheduling requests to minimize energy consumption. &\khan{\xmark} &\khan{\checkmark} &\khan{\checkmark} &\khan{\xmark} &\khan{\xmark} &\khan{\xmark} &\khan{\xmark} \\\cline{2-10}

& \cite{tran_2017} &Jointly optimize caching and transcoding resources using ILP to minimize the backhaul network cost. &\khan{\checkmark} &\khan{\xmark} &\khan{\checkmark} &\khan{\checkmark} &\khan{\xmark} &\khan{\xmark} &\khan{\xmark} \\\cline{2-10}
&\cite{Jiang2019} & Cooperative caching when users' preference is unknown and only the historical content demands. &\khan{\checkmark} &\khan{\xmark} &\khan{\checkmark} &\khan{\xmark} &\khan{\xmark} &\khan{\xmark} &\khan{\xmark} \\\cline{2-10}
&\cite{Wang2019b} & Joint caching placement (long-term) and video retrieval (short-term) in coordinated multi-server system. &\khan{\checkmark} &\khan{\xmark} &\khan{\checkmark} &\khan{\xmark} &\khan{\xmark} &\khan{\xmark} &\khan{\xmark} \\\cline{2-10}
&\cite{Chien2020a} & Using Q-learning, find the appropriate cache state to improve caching mechanism. &\khan{\xmark} &\khan{\xmark} &\khan{\xmark} &\khan{\xmark} &\khan{\xmark} &\khan{\xmark} &\khan{\checkmark} \\\cline{2-10}
&\cite{Tran2019} & Joint collaborative caching and processing at multiple servers to minimize the delay of video retrieval. &\khan{\checkmark} &\khan{\xmark} &\khan{\checkmark} &\khan{\checkmark} &\khan{\xmark} &\khan{\xmark} &\khan{\xmark} \\ \cline{2-10}
&\cite{Baccour2020} & Cache only the video chunks to be watched and collaboration among neighboring MECs to improve utilization. &\khan{\checkmark} &\khan{\xmark} &\khan{\checkmark} &\khan{\checkmark} &\khan{\xmark} &\khan{\xmark} &\khan{\xmark} \\\cline{2-10}
&\khan{\cite{sang2021}} & \khan{A greedy collaborative caching strategy in heterogenous MEC network to minimize total delay of all users.}. &\khan{\checkmark} &\khan{\xmark} &\khan{\checkmark} &\khan{\checkmark} &\khan{\xmark} &\khan{\xmark} &\khan{\xmark} 
\\[0.8em]  \hline \hline

\multirow{2}{2cm}{Cooperative Processing} 
&\cite{Sun2020} &Edge nodes pre-process video data using DNN model and upload the results to cloud  for further analysis. &\khan{\checkmark} &\khan{\checkmark} &\khan{\xmark} &\khan{\xmark} &\khan{\xmark} &\khan{\xmark} &\khan{\xmark} \\\cline{2-10}
&\khan{\cite{taha2021}} &\khan{Peer-offloading using perceptual parameters such as pausing frequency, watching percentage and bit rates.} &\khan{\xmark} &\khan{\xmark} &\khan{\xmark} &\khan{\xmark} &\khan{\xmark} &\khan{\xmark} &\khan{\checkmark}   \\[0.8em]  \hline \hline

\multirow{16}{2cm}{Joint Caching and Processing} 
&\cite{Yang2018a} &Reduce communication-resource consumption by using MEC to compute components not stored on VR device. &\khan{\xmark} &\khan{\xmark} &\khan{\xmark} &\khan{\checkmark} &\khan{\xmark} &\khan{\xmark} &\khan{\xmark}  \\\cline{2-10}
&\cite{Yan2019} &Implement per-channel optimal service-aware resource allocation. &\khan{\xmark} &\khan{\xmark} &\khan{\xmark} &\khan{\xmark} &\khan{\xmark} &\khan{\xmark} &\khan{\checkmark} \\\cline{2-10}
&\cite{Tran2019} &A method to decide (i) transcode locally or (ii) fetch from neighboring MEC or (iii) fetch from the origin server. &\khan{\checkmark} &\khan{\xmark} &\khan{\checkmark} &\khan{\checkmark} &\khan{\xmark} &\khan{\xmark} &\khan{\xmark} \\\cline{2-10}
&\cite{Hao2019} &A video caching and processing model that offers maximized profit to video service provider. &\khan{\xmark} &\khan{\xmark} &\khan{\xmark} &\khan{\xmark} &\khan{\xmark} &\khan{\checkmark} &\khan{\xmark} \\\cline{2-10}
&\cite{Xie2019} &A joint caching and transcoding scheduling strategy to minimize the energy consumption. &\khan{\xmark} &\khan{\checkmark} &\khan{\xmark} &\khan{\xmark} &\khan{\xmark} &\khan{\xmark} &\khan{\xmark} \\\cline{2-10}
&\cite{Tang2019} &BSs compete to maximise its revenue by maximising the request rate that can be served by the MEC server. &\khan{\xmark} &\khan{\xmark} &\khan{\xmark} &\khan{\xmark} &\khan{\xmark} &\khan{\xmark} &\khan{\checkmark} \\\cline{2-10}
&\cite{Zhang2020b} &Minimizes the average retrieval latency of all users using proactively caching and user-BS association scheme. &\khan{\checkmark} &\khan{\xmark} &\khan{\checkmark} &\khan{\checkmark} &\khan{\xmark} &\khan{\xmark} &\khan{\xmark} \\\cline{2-10}
&\cite{Baccour2020} & Proactively cache video chunks likely to be watched instead of whole video content. &\khan{\checkmark} &\khan{\xmark} &\khan{\checkmark} &\khan{\checkmark} &\khan{\xmark} &\khan{\xmark} &\khan{\xmark} \\\cline{2-10}
&\cite{Bilal2019} &Online bit rate conversion to the requested version of videos fetched from the origin/CDN servers. &\khan{\checkmark} &\khan{\xmark} &\khan{\checkmark} &\khan{\checkmark} &\khan{\xmark} &\khan{\xmark} &\khan{\xmark} \\\cline{2-10}
&\cite{Zhang2020} &Joint task offloading and dynamic caching strategy to reduce overall latency of all mobile devices.  &\khan{\checkmark} &\khan{\checkmark} &\khan{\xmark} &\khan{\xmark} &\khan{\xmark} &\khan{\xmark} &\khan{\xmark} \\\cline{2-10}
&\khan{\cite{huang2021}} & \khan{Joint optimization of caching and transcoding for resource allocation in DASH.} &\khan{\xmark} &\khan{\xmark} &\khan{\checkmark} &\khan{\xmark} &\khan{\checkmark} &\khan{\xmark} &\khan{\checkmark} \\[0.8em]  \hline \hline

\multirow{10}{2cm}{Joint Radio, Caching and Processing} 
&\cite{Dang2019} &To maximize the average tolerant delay while meeting the rate constraint in VR devices. &\khan{\checkmark} &\khan{\xmark} &\khan{\xmark} &\khan{\xmark} &\khan{\xmark} &\khan{\xmark} &\khan{\xmark}  \\\cline{2-10}
&\cite{Sun2019} &To decide whether to pre-cache (and if yes, which) parts of the field of views (FOVs), with local pre-processing. &\khan{\xmark} &\khan{\xmark} &\khan{\xmark} &\khan{\checkmark} &\khan{\xmark} &\khan{\xmark} &\khan{\xmark}  \\\cline{2-10}
&\cite{Luo2019} &Jointly considers buffers, video quality, edge caching, transcoding and transmission for energy saving and QoE. &\khan{\xmark} &\khan{\checkmark} &\khan{\xmark} &\khan{\xmark} &\khan{\xmark} &\khan{\xmark} &\khan{\xmark} \\\cline{2-10}
&\cite{Wang2020h} &Jointly optimize the caching, radio and resource allocation to maximize the system revenue. &\khan{\xmark} &\khan{\xmark} &\khan{\xmark} &\khan{\xmark} &\khan{\xmark} &\khan{\checkmark} &\khan{\xmark} \\\cline{2-10}
&\cite{Rezvani2018} & Joint multi-bitrate caching and transcoding by allocating physical and radio resources based on network stats.  &\khan{\xmark} &\khan{\xmark} &\khan{\xmark} &\khan{\xmark} &\khan{\xmark} &\khan{\checkmark} &\khan{\xmark} \\\cline{2-10}
&\khan{\cite{cheng2021}} & \khan{Jointly considers transmission, coding, caching, and computation to meet delay requirement in mobile VR.}  &\khan{\checkmark} &\khan{\xmark} &\khan{\xmark} &\khan{\xmark} &\khan{\xmark} &\khan{\xmark} &\khan{\xmark} \\\cline{2-10}
&\khan{\cite{liu2021}} & \khan{Jointly optimize cachig, processing and transmission to reduce latency}. &\khan{\checkmark} &\khan{\xmark} &\khan{\xmark} &\khan{\xmark} &\khan{\xmark} &\khan{\xmark} &\khan{\xmark}  \\[0.8em]  \hline




\end{tabular}

%% file: tables/d2d_mec.tex
\begin{tabular}{|m{1.6cm} | >{\arraybackslash}m{0.5cm} | >{\arraybackslash}m{2.6cm} | >{\arraybackslash}m{1.3cm} | >{\arraybackslash}m{2.2cm} | >{\arraybackslash}m{7cm} |} \hline

\textbf{Category} &\textbf{Ref} &\textbf{Algorithm} &\textbf{\khan{Evaluation Method}} &\textbf{\khan{Metric}} &\textbf{Description}  \\[0.5em] \hline \hline

\multirow{3}{2cm}{Methods and Use cases}
&\cite{Li2014} &Numerical model &\khan{Not Available}  &\khan{Task success rate}
&Propose (i) optimal and (ii) periodic scheme to offload tasks and retrieve results. \\[0.5em]

&\cite{Pu2016} &Lyapunov optimization &\khan{Simulation} &\khan{Energy}
&Dynamic D2D offloading that minimize energy consumption of all users, while preventing over-exploiting helper device's resources \\[1em]

&\cite{He2019} &MIP &\khan{Simulation} &\khan{Capacity}
&Maximize the number of devices supported by cellular network by minimizing edge resource computation and optimal D2D pairing. \\[1em]

&\cite{Yuan2019} &Lagrangian Multipliers &\khan{Simulation} &\khan{Capacity}
&Propose a file allocation algorithm in D2D MEC systems to improve the caching capacity  \\[1em]


&\cite{wang_2017b}  &Convex optimization &\khan{Simulation} &\khan{Energy}
&Collaborative beamforming to wirelessly charge other devices using D2D and offloading computational tasks to idle devices. \\[1em]

&\cite{xing_2019}  &MINLP &\khan{Simulation} &\khan{Latency, capacity}
&User offload computational tasks to multiple local users using time division multiple access (TDMA). \\[1em] \hline \hline

\multirow{3}{2cm}{Resource Optimizing}
&\cite{Wu2019} &Convex optimization  &\khan{Simulation} &\khan{Capacity}
&Jointly optimize energy beam-forming, communication and computation resources to maximize sum of computation rates of users. \\[1em]

& \cite{Qiao2019} &Q Learning  &\khan{Simulation} &\khan{Latency}
&Jointly optimize computational offloading, power allocation and CPU frequency adjustment for computation offloading and resource management. \\[1em]

&\cite{Kim2020}  &NLIP &\khan{Simulation} &\khan{Energy}
&Select links in D2D to minimize energy consumption and then jointly optimize resource allocation and MIMO signal design. \\[1em] \hline \hline

\multirow{2}{2cm}{Performance Evaluation}
&\cite{dogga_2019}  &Android testbed  &\khan{Real data} &\khan{Capacity, Energy}
&Transcoded videos of short duration on mobile phones to measure the transcoding time. \\[1.5em]

\hline
\end{tabular}

%% file: tables/mec_ai.tex

\begin{tabular}{ |>{\arraybackslash}m{2cm} | >{\centering\arraybackslash}m{0.5cm} | >{\centering\arraybackslash}m{2cm} | >{\centering\arraybackslash}m{1.2cm} | >{\centering\arraybackslash}m{1.2cm} | >{\arraybackslash}m{8cm}|}

\hline
\textbf{Category} &\textbf{Ref} &\textbf{Model} &\khan{\textbf{Method}} & \khan{\textbf{Dataset}} &\textbf{Description} \\[0.5em]
\hline \hline

& \cite{li2013popularity} &ARIMA, MLR, KNN &\khan{Testbed} &\khan{Real}
&Forecast video popularity pattern to cache popular videos \\[0.5em]
& \cite{wang2017edge}  &Q-learning  &\khan{Simulation} &\khan{Synthetic}
& Distributed cache replacement using Q-learning based on content popularity \\[0.5em]
&\cite{tanzil2017adaptive}  &ELM  &\khan{Simulation} &\khan{Real}
& Estimate the unknown popularity of contents for caching. \\[0.5em]
& \cite{zhu2018deep} &DRL  &\khan{Simulation} &\khan{Real}
&Learn end-to-end caching policy using DRL \\[0.5em] 
& \cite{Tan2018} &DQN  &\khan{Simulation} &\khan{Synthetic}
&Estimating mobility-aware reward for joint caching and computing in edge.
\\ [0.5em]
& \cite{zhang2019toward} &LSTM  &\khan{Simulation} &\khan{Real}
&Cache videos using long-term and short-term popularity prediction of videos. \\ 
\multirow{-4}{*}{Caching} &&&&& \\[1em]\hline\hline

& \cite{wei2018joint} &DRL  &\khan{Simulation} &\khan{Synthetic}
&Minimizes end-to-end delay for caching, offloading, and radio resources \\
& \cite{huang2019deep} &DQN  &\khan{Simulation} &\khan{Synthetic}
&Minimizes energy, computation, and delay cost for multiple task offloading \\
\multirow{-5}{2cm}{Computation Offloading} &&&&& \\[1em]\hline\hline

& \cite{de2017qoe} &RBM  &\khan{Simulation} &\khan{Real}
&Using RBM with linear classifier for concurrent transmission to multiple users with QoS guarantees\\
& \cite{wang2019intelligent} &DRL  &\khan{Simulation} &\khan{Real}
&Use DRL to assign users to the appropriate server in multi server MEC system. \\ 
& \khan{\cite{yang2021}} &\khan{DRL}  &\khan{Simulation} &\khan{Synthetic}
&\khan{MEC-assisted adaptive video streaming for driving assistance for improved "quality level switching".} \\ 
\multirow{-6}{2cm}{Adaptive Video Streaming \& QoS} &&&&& \\[1em]\hline\hline

& \cite{ren2018distributed} &CNN  &\khan{Testbed} &\khan{Real}
& Edge-based faster object detection in videos. \\
& \cite{liu2017new} &CNN  &\khan{Testbed} &\khan{Real}
&Edge-based object recognition model for food items.  \\
& \cite{ran2018deepdecision} &CNN  &\khan{Testbed} &\khan{Real}
& Improving frame rate and accuracy with edge computing running CNN. \\ 
& \cite{li2016deepcham} &CNN  &\khan{Simulation} &\khan{Real}
&Collaboration among user and edge server to train a CNN Model to improve object recognition. \\ 
& \cite{nikouei2018smart} &L-CNN  &\khan{Simulation} &\khan{Real}
&A CNN model for resource-constrained edge server.\\ 
& \cite{Wang2020c} &ADMM  &\khan{Simulation} &\khan{Real}
&Latency-aware video summarization using ADMM, to understand the \khan{story-line} of a video before a client requests the complete video content summarizing\\ [0.5em]
& \khan{\cite{galano2021}} &\khan{DNN}  &\khan{Testbed} &\khan{Real}
&\khan{Automated ML framework using DNN to perform object recognition in device-captured frames sent at the edge server, while improving recognition accuracy.}\\
\multirow{-8}{2cm}{Video Analytics} &&&&& \\ \hline

\end{tabular}

%% file: tables/video_datasets.tex
\begin{tabular}{ |>{\arraybackslash}m{2.5cm} | >{\centering\arraybackslash}m{0.5cm} | >{\centering\arraybackslash}m{0.5cm} | >{\arraybackslash}m{9cm} | >{\arraybackslash}m{7cm} |}
\hline


    \textbf{Dataset}  &\textbf{Ref}  &\textbf{Year}  &\textbf{Description} &\textbf{Applications}  \\[2em] \hline \hline

     Facebook Live    &\cite{facebooklive_2018}     &2018
     &The dataset consists of  a list of 1.506.473 videos collected from Facebook live over a period of 34 days. The videos are published by 408.231 different broadcasters.
     & Multi-cloud resource allocation, edge caching, and traffic allocation problem.
     \\[2em]\hline
     
     YouTube-8M&\cite{youtube8m}  &2019
     & The original dataset contains 6.1 Million video IDs, 350,000 hours of videos and 3862 classes. The average number of machine-generated labels for each video are 3. The extended version of the dataset contains human-verified labels on about 237K segments (average of 5 segments/video) on 1000 classes from the validation set.
     &Video analytics such as activity recognition, captioning, and classification.
     \\[3em]\hline

     Twitch dataset  &\cite{twitch_dataset}  &2015
     & A dataset of crawled data collected from Twitch every five minutes in a one-month period. The dataset consists of 11 parameters e.g. number of viewers, number of active viewers, followers gained and parameters of particular streamer.
     &User-generated live streaming
     \\[2em]\hline

     VR Streaming dataset&\cite{vr_dataset}  &2017
     &  The dataset contains head tracking of 48 users (24 males and 24 females) watching 18 spherical videos from 5 categories.
     & User behavior patterns in VR streaming, visual attention modeling, gazing prediction, user identification based on head motion pattern.
     \\[2em]\hline

     $360^{\circ}$ Videos dataset &\cite{360videos}  &2019
     & A dataset of 28 $360^{\circ}$ videos based on the taxonomy, and recorded viewport traces from 60 participants watching the videos.
     &Viewing experience in $360^{\circ}$ video streaming 
     \\[2em]\hline

     LIVE Mobile Stall Video Database-II &\cite{video_stall}  &2017
     & The dataset contains a total of 174 distorted videos generated from 24 reference videos with 26 unique stalling events. It also includes both continuous-time and overall subjective scores from 54 unique subjects.
     & QoE predictive models, performance benchmarking, and user-centric mobile network planning
     \\[4em]\hline

     YouTube's mobile video streaming dataset &\cite{youtube_mobile_dataset}  &2019
     & Dataset is acquired for 783.88 hours of playback time and contain several network and traffic related parameters i.e. type of measurement, timestamp, scenario, videoID, packet arrival timestamp, packet length in bytes, source and destination IP address, source and destination TCP port and the payload protocol number, TCP flow duration, flow size and initial video playback delay.
     & Studies on mobile videos initial delay.
     \\[4em]\hline

     5G trace dataset &\cite{5G_video_dataset}  &2020
     & The dataset is generated from two static and moving cars, using two different applications i.e. video streaming and file download. The dataset contains client-side cellular key performance indicators (KPIs) e.g. channel-related metrics, context-related metrics, cell-related metrics and throughput information.
     & Video streaming QoE in 5G networks
     \\[3em]\hline

     UVG dataset &\cite{uvg_dataset}  &2020
     & Dataset for Ultra Video Group (UVG) video codec analysis, consisting of 16 versatile 4K video sequences, captured either at 50 or 120 frames per second (fps).
     & Subjective and objective quality assessments of next-generation VVC codecs.
     \\[2em]\hline
     
     LIVE-NFLX-II &\cite{LIVE_NFLX_II}  &2018
     & A subjective Video QoE dataset which includes 420 videos that were evaluated by 65 subjects. These videos were generated from 15 video contents streamed under 7 different network conditions and 4 client adaptation strategies.
     & QoE studies e.g. video quality fluctuations, rebuffering events of varying durations and numbers, spatial resolution changes, and diverse bitrate/quality levels and video content types.
     \\[3em] \hline

     \hline
    \end{tabular}